\documentclass[final,3p,times]{elsarticle}

\usepackage{graphicx}
\usepackage{amsmath}
\usepackage{siunitx}
\usepackage{longtable,tabularx}
\usepackage{physics}
\usepackage{amssymb}
\usepackage{physics}
\usepackage{graphicx}
\usepackage{mathrsfs}
\usepackage{color}
\usepackage{diagbox}
\usepackage{subcaption}
\usepackage{arydshln}

\makeatletter
\def\ps@pprintTitle{%
 \let\@oddhead\@empty
 \let\@evenhead\@empty
 \def\@oddfoot{}%
 \let\@evenfoot\@oddfoot}
\makeatother

\begin{document}

\begin{frontmatter}

\title{RBF-FD discretization of the Navier-Stokes equations on scattered but staggered nodes}



\author{Tianyi Chu}

        
            \ead{tic173@eng.ucsd.edu}

\author{Oliver T. Schmidt\corref{cor1}}
            \ead{oschmidt@eng.ucsd.edu}
            
          \address{Department of Mechanical and Aerospace Engineering, Jacobs School of Engineering, UCSD, 9500 Gilman Drive, La Jolla, CA92093, USA.}

 \cortext[cor1]{Corresponding author}

\begin{abstract}

A semi-implicit fractional-step method that uses a staggered node layout  and radial basis function-finite differences (RBF-FD) 
to solve the incompressible Navier-Stokes equations is developed. Polyharmonic splines (PHS) with polynomial augmentation (PHS+poly) are used to construct the global differentiation matrices. A systematic parameter study identifies a combination of stencil size, PHS exponent, and polynomial degree that minimizes the truncation error for a wave-like test function on scattered nodes. Classical modified wavenumber analysis is extended to RBF-FDs on heterogeneous node distributions and used to confirm that the accuracy of the selected 28-point stencil is comparable to that of spectral-like, 6th-order Pad\'e-type finite differences. The Navier-Stokes solver is demonstrated on two benchmark problems, internal flow in a lid-driven cavity in the Reynolds number regime $10^2 \leq \mathrm{Re}\leq 10^4$, and open flow around a cylinder at $\mathrm{Re}= 100$ and 200. The combination of grid staggering and careful parameter selection facilitates accurate and stable simulations at significantly lower resolutions than previously reported, using more compact RBF-FD stencils, without special treatment near solid walls, and without the need for hyperviscosity or other means of regularization.


\end{abstract}


\end{frontmatter}

\section{Introduction}

In the past two decades, radial basis functions (RBF)-based discretizations have emerged as a viable alternative to established approaches for computational fluid mechanics \citep{fornberg2015solving}. RBF-based methods are often referred to as \emph{mesh-free} as they facilitate the discretization of partial differential operators directly on a set of scattered nodes, i.e., without the need of local elements. The main promise of RBF-based methods is that they can combine the ease of implementation and high order of accuracy of finite differences (FD) with the geometrical flexibility of finite volume (FV), element, and discontinuous Galerkin (DG) methods.

 RBF approximations provide high-order accuracy, flexibility, and ease of implementation for interpolation and differentiation.
    Historically, RBFs were often used as global interpolants over all nodes. These global RBF methods with spectral-like nominal accuracy have been applied to a variety of flow problems \citep{flyer2011radial,flyer2010rotational,flyer2007transport,flyer2009radial,kansa1990multiquadrics1,kansa1990multiquadrics2,wright2010hybrid} including Lagrangian fluid mechanics \citep{wang2020weighted}. 
Other applications of global RBFs include solid mechanics \citep{wang2017radial, wang2021static}.
They are, however, computationally expensive for large problems and often suffer from ill-conditioning and numerical instability. Remedies to these problems have been found in the form of regularizations like hyperviscosity \citep{flyer2012guide,fornberg2011stabilization} and preconditioners \citep{kansa2000circumventing}.
    The use of local RBF stencils was pioneered by \citep{shu2003local,tolstykh2000using,wang2002point,wright2003radial} and yields a class of so-called RBF-FD methods that are named in reference to classical finite differences. Just like classical FDs, RBF-FD methods generate sparse differentiation matrices. The level of sparsity depends on the local stencil size, which in turn is determined by the desired order of accuracy. Common choices of RBFs for fluid flow problems are Gaussians (GA), multiquadrics (MQ), and inverse multiquadrics (IMQ).
\citet{flyer2012guide}, for example, demonstrated the use of GA-type RBFs for solving the shallow water equations on a sphere and compared the performance to other high-order numerical methods.
 Applications to the incompressible Navier-Stokes equations (NSE) include the local MQ-differential quadrature (DQ) method by \citet{shu2005computation}, its extension to 3D by \citet{ding2006numerical}, the compact RBF-FD scheme by \citet{chinchapatnam2009compact}, which in turn is based on the work by \citet{wright2006scattered}, and the method by \citet{xie2021improved}, who introduced a regularization term for IMQ RBFs and a semi-Lagrangian scheme for transient simulations. Incompressible flow solvers with convective heat transfer have been implemented by \citep{shu2003local,waters2015global,zamolo2019solution}.

All these implementations use infinitely smooth RBFs that are characterized by a shape parameter.
This shape parameter, in turn, significantly impacts both accuracy and stability, and extensive works, most empirical, have been devoted to investigating its effect \citep{carlson1991parameter,fasshauer2007choosing,fornberg2004stable,franke1982scattered,hardy1971multiquadric,rippa1999algorithm}.
Good accuracy is often associated with near-flat RBFs. RBFs in this flat limit, however, often yield ill-conditioned discretizations and suffer from  stagnation, or saturation, errors \citep{flyer2016role}.
Numerical schemes that address these problems include Contour-Pad\'e \citep{fornberg2004stable}, RBF-QR \citep{fornberg2011stable,fornberg2008stable,larsson2013stable}, and RBF-GA \citep{fornberg2013stable} methods.
More recently, RBF-FDs based on polyharmonic splines augmented with polynomials (PHS+poly), that do not require a shape parameter were introduced by \citet{flyer2016enhancing}.
Later, \citet{flyer2016role} demonstrated the use of higher-order polynomial augmentations, which improve the accuracy of derivative approximations of local RBF stencils and mitigate the stagnation error under node refinement.
\citet{bayona2017role} used PHS+poly for solving elliptic PDEs and showed that a larger stencil size near domain boundaries helps to avoid the Runge phenomenon. Using closed-form RBFs, \citet{bayona2019insight} later provided an analytical validation of this result. Numerical demonstrations for 2-D and 3-D examples were presented in \citet{bayona2019role}. 
Several comparisons to other mesh-free approaches, including polynomial least-squares approximations \citep{flyer2016role}, the RBF-GA method near the flat limit \citep{santos2018comparing}, and the moving least-squares (MLS) method \citep{bayona2019comparison} have confirmed the competitiveness of PHS+poly-based RBF-FDs in terms of accuracy, robustness, and computational efficiency.
While there are numerous demonstrations of PHS+poly RBF-FDs for advection-diffusion problems \citep{bartwal2021application,gunderman2020transport,shankar2017overlapped,shankar2018hyperviscosity,shankar2018rbf,su2019radial},
their application to the incompressible Navier-Stokes equations has been explored only very recently. \citet{shahane2021high}, for example, simulated incompressible flows using an explicit fractional-step method and considered several test problems to gauge accuracy. A semi-implicit algorithm was proposed by \citet{shahane2021semi} and later implemented by \citet{unnikrishnan2021shear} to simulate Taylor-Couette flow.
The consistency and convergence of this algorithm with respect to grid resolution were later examined in \citet{shahane2022consistency}.

 We build on this successful combination of fractional-step methods with PHS+poly RBF-FD discretizations and make two main contributions, one in terms of improving computational efficiency and the other in terms of error analysis. 
 A significant reduction in computational cost is achieved through the use of a staggered-grid arrangement that permits the use of smaller stencils and, at the same time, much coarser grids.
The staggered grid arranges the velocity and pressure at different nodes to circumvent the numerical instability known as odd-even decoupling. 
While the idea of grid staggering originates from the classical FD Marker-And-Cell scheme by \citet{harlow1965numerical}, staggering occurs naturally in finite volume methods that define velocity in terms of fluxes across cell faces. 
    Similarly, the proposed staggering strategy is based on an underlying triangular grid, generated using any standard grid generator, and defines the pressure at the vertices and velocities at the centers of the faces.

The second contribution is the use of modified wavenumber diagrams, as known from classical FD analysis, to gauge the accuracy of the PHS+poly RBF-FDs on the staggered nodes.
    Modified wavenumber analysis was originally designed to examine the truncation error of FD methods on lattice-based grids but has also been applied to unstructured FD discretizations  \citep{nishikawa2021flexible,park2007numerical}.
 Accuracy and error analysis for RBF approximations are not straightforward and often done in a problem-specific manner.
   For PHS+poly RBF-FDs, for example, test problems such as the summation of sinusoids \citep{flyer2016enhancing,flyer2016role}, Poisson’s equation \citep{bayona2017role}, Kovasznay flow, and cylindrical Couette flow \citep{shahane2021high} were considered.
    These studies mainly focused on the discretization error under grid refinement and have demonstrated that the convergence rate is determined by the degree of the polynomial augmentation.
    Here, we first conduct a systematic parameter study to identify a set of parameters, i.e., stencil size, PHS exponent, and polynomial degree, that minimizes the relative error for a wave-like test function on a representative, highly heterogeneous grid.
We proceed with a two-dimensional modified wavenumber analysis to quantify the order of accuracy on the staggered grids. The analysis shows that the selected 28-point stencil PHS+poly RBF-FDs provide accuracy comparable to 6th-order Pad\'e-type FDs.

This paper is organized as follows:
\S \ref{RBF} introduces the PHS+poly RBF-FD method,
\S \ref{Spatial Discretization} describes the unstructured staggered node arrangement and the fractional-step method, and
\S \ref{error_analysis} the parameter selection and error and accuracy analysis for the staggered grids.
  The performance of the incompressible Navier-Stokes solver is demonstrated in \S \ref{applications} on the lid-driven cavity and cylinder flow as benchmark problems. 
Finally, \S \ref{conclusion} concludes and summarizes the paper.

\section{Radial basis functions (RBFs)}  \label{RBF}

The underlying idea of radial basis functions (RBFs) is to approximate a given function $f(\vb*{x})$ using a
 smooth radial function $\phi(r)$. 
 For a set of $n$ scattered nodes, $\{\vb*{x}\}_{j=1}^n$, we seek the interpolant 
\begin{equation}
    s(\vb*{x})=\sum_{j=1}^n \lambda_j \phi(\|\vb*{x}-\vb*{x}_j\|) \label{rbf_intp}
\end{equation}
that satisfies $s(\vb*{x}_i)=f(\vb*{x}_i)$ for $i= 1,2,\dots ,n$, where $\|\cdot\|$ denotes the
standard Euclidean norm. 
The interpolation coefficients $\lambda_1,\cdots,\lambda_n$ can be found as the solution of the linear system
\begin{equation}
   \underbrace{\mqty[\phi(\|\vb*{x}_1-\vb*{x}_1\|) & \phi(\|\vb*{x}_1-\vb*{x}_2\|) & \cdots & \phi(\|\vb*{x}_1-\vb*{x}_n\|)\\
    \phi(\|\vb*{x}_2-\vb*{x}_1\|) & \phi(\|\vb*{x}_2-\vb*{x}_2\|) & \cdots & \phi(\|\vb*{x}_2-\vb*{x}_n\|)\\
    \vdots & \vdots &  & \vdots\\
       \phi(\|\vb*{x}_n-\vb*{x}_1\|) & \phi(\|\vb*{x}_n-\vb*{x}_2\|) & \cdots & \phi(\|\vb*{x}_n-\vb*{x}_n\|)
    ]}_{\vb*{A}} \mqty[ \lambda_1 \\ \lambda_2 \\ \vdots \\ \lambda_n]=\mqty[ f(\vb*{x}_1) \\ f(\vb*{x}_2) \\ \vdots \\ f(\vb*{x}_n)], \label{A}
\end{equation}
where $\vb*{A}$ is the interpolation matrix.
The RBF interpolant $s(\vb*{x})$ based on these coefficients can then be used to approximate the function $f(\vb*{x})$ in the local region described by the set of nodes, $\{\vb*{x}\}_{j=1}^n$.
Common choices for RBFs include Gaussian (GA), multiquadrics (MQ), and inverse multiquadrics (IMQ). See \cite{fornberg2015solving} for a comprehensive overview.
These RBF types are known to suffer from the stagnation (or saturation) error under refinement and have a free shape parameter that significantly impacts their accuracy and stability.
In this work, we use polyharmonic splines (PHS),
\begin{align}
    \phi(r) =r^m,\qquad \text{where}\,\, m\,\, \text{is an odd positive integer,}
\end{align}
as the basis functions. This choice is motivated by recent studies  \cite{bayona2017role,bayona2019role,flyer2016enhancing,flyer2016role},
which highlight the advantageous properties of the PHS-type RBFs for discretizations, described next.

\subsection{RBF-FD method and augmentation with polynomials}

A direct approach for the generation of RBF-based differentiation operations is the RBF-FD method, which 
approximates the action of any linear operator, $\mathcal{L}$, as a linear combination of the function values, $f(\vb*{x}_j)$, such that 
\begin{equation}
   \mathcal{L}f(\vb*{x}_0)=\sum_{j=1}^n w_{j} f(\vb*{x}_j). \label{Lf} 
\end{equation}
Here, $x_0$ is a given location, and $w_j$ are the unknown differentiation weights.
Using equation (\ref{rbf_intp}), the function value $f(\vb*{x})$ can be approximated by the RBF interpolant $s(\vb*{x})$ and the corresponding weight vector $\vb*{w}=(w_1,\cdots, w_n)^T$ is then obtained by solving the linear system
\begin{equation}
   \mqty[& &\\&\vb*{A}&\\ & &] \mqty[ w_{ 1}  \\ w_{ 2} \\ \vdots  \\ w_{ n} ]=\mqty[ \eval{\mathcal{L}\phi(\|\vb*{x}-\vb*{x}_1\|)}_{\vb*{x}=\vb*{x}_0} \\
   \eval{\mathcal{L}\phi(\|\vb*{x}-\vb*{x}_2\|)}_{\vb*{x}=\vb*{x}_0}  \\
   \vdots \\
   \eval{\mathcal{L}\phi(\|\vb*{x}-\vb*{x}_n\|)}_{\vb*{x}=\vb*{x}_0}]. \label{w_local}
\end{equation}
An implicit assumption of the RBF-FD method is that the derivative of the basis functions, $\mathcal{L}\phi$, is continuous.

A commonly used extension of equation (\ref{Lf}) to enforce consistency with Taylor expansion-based FD approximations is polynomial augmentation \cite{flyer2016enhancing,fornberg2011stabilization,fornberg2015solving,larsson2013stable,wright2006scattered}.
For 2D problems, this RBF-FD method with polynomial augmentation up to degree $q$ takes the form
\begin{equation}
   \mathcal{L}f(\vb*{x}_0)=\sum_{j=1}^n w_{j} f(\vb*{x}_j) + \sum_{i=1}^{(q+1)(q+2)/2} c_i P_i(\vb*{x}_0). \label{Lf_PHS} 
\end{equation}
The use of multivariate polynomial terms, $P_i(\vb*{x})$, to match the local Taylor series introduces the additional constraints
\begin{align}
    \sum_{j=1}^n w_j P_i(\vb*{x}_j) =  \mathcal{L} P_i(\vb*{x}_0), \qquad \text{for} \,\, 1\leq i\leq \frac{(q+1)(q+2)}{2},
\end{align}
also known as the vanishing momentum conditions \cite{iske2003approximation}, for the differentiation weights.
These constraints ensure that the RBF approximations reproduce locally polynomial behaviour up to degree $q$ \cite{flyer2016role} and 
appropriately decay in the far-field \cite{fornberg2002observations}.
As an example, consider the linear system for $q=1$,
\begin{equation} \label{rbf_poly}
   \left[
\begin{array}{c c  c; {2pt/2pt} c c c}
 & & & 1 & x_1 & y_1\\
    &\vb*{A} & & \vdots & \vdots & \vdots \\ 
    & &  & 1 & x_n & y_n  \\ \hdashline[2pt/2pt]
    1 & \cdots & 1 & & & \\
    x_1 &  \cdots  & x_n & &\vb*{0} &\\
    y_1 &  \cdots  & y_n& & & 
   \end{array}\right]
   \mqty[ w_{ 1}   \\ \vdots  \\ w_{ n} \\\hdashline[2pt/2pt] c_1 \\c_2\\c_3]=\mqty[ \eval{\mathcal{L}\phi(\|\vb*{x}-\vb*{x}_1\|)}_{\vb*{x}=\vb*{x}_0} \\
   \vdots \\
   \eval{\mathcal{L}\phi(\|\vb*{x}-\vb*{x}_n\|)}_{\vb*{x}=\vb*{x}_0}\\ \hdashline[2pt/2pt]
   \eval{\mathcal{L}1}_{\vb*{x}=\vb*{x}_0}\\
   \eval{\mathcal{L}x}_{\vb*{x}=\vb*{x}_0}\\
    \eval{\mathcal{L}y}_{\vb*{x}=\vb*{x}_0}]. 
\end{equation}
The interpolation matrix $\vb*{A}$ is the same one defined in equation (\ref{w_local}). A more general and compact expression for equation (\ref{rbf_poly}) takes the form
\begin{equation} \label{rbf_poly_compact}
 \underbrace{  \left[
\begin{array}{c  c }
    \vb*{A} &  \vb*{P} \\ 
   \vb*{P}^{\small{T}} & \vb*{0}
   \end{array}\right]}_{\vb*{A}_\text{aug}}
   \mqty[\vb*{w}\\ \vb*{c}]=\mqty[{\mathcal{L}\vb*{\phi}}\\
   {\mathcal{L}\vb*{P}}]. 
\end{equation}
This procedure ensures consistency with the local expansion, but equation (\ref{Lf}) is used to approximate the actual differentiation operations.
The use of polynomial augmentation for PHS RBFs has shown to improve the accuracy for interpolation and derivative approximations \cite{flyer2016role}, numerical solutions of elliptic PDE problems \cite{bayona2017role}, and approximations near domain boundaries \cite{bayona2019role}. Our practical implementation follows \citep{flyer2016enhancing}.

\subsection{Global differentiation operators}

Differentiation matrices provide a straightforward and flexible way to discretize partial differential equations (PDEs).
The use of global RBFs leads to full matrices, which is it is computationally expensive and requires a lot of memory. 
To obtain sparse matrices instead, we seek local differentiation operators that utilize a smaller number of neighboring nodes.
Assume the given domain is discretized by two sets of scattered nodes, $\{\vb*{x}_i^{(\alpha)}\}_{i=1}^N$ and $\{\vb*{x}_j^{(\beta)}\}_{j=1}^M$.
Given the function values at node set $\beta$, $f(\vb*{x}^{(\beta)})$, we seek a differentiation matrix such that $\vb*{D}_{\mathcal{L}}^{(\alpha,\beta)} f(\vb*{x}^{(\beta)})$ approximates the derivatives, $\mathcal{L}f(\vb*{x}^{(\alpha)})$, at node set $\alpha$.
The differentiation matrix $\vb*{D}_{\mathcal{L}}^{(\alpha,\beta)} $ must hence satisfy
\begin{equation}
   \underbrace{\mqty[w_{11} & w_{12} & \cdots & w_{1N}\\
    w_{21} & w_{2 2} & \cdots & w_{2 N}\\
    \vdots & \vdots &  & \vdots\\
       w_{M 1} & w_{M 2} & \cdots & w_{M N}
    ]}_{\vb*{D}_{\mathcal{L}}^{(\alpha,\beta)} } \mqty[ f(\vb*{x}_1^{(\beta)}) \\ f(\vb*{x}_2^{(\beta)}) \\ \vdots \\ f(\vb*{x}_N^{(\beta)})]=\mqty[ \mathcal{L} f(\vb*{x}_1^{(\alpha)}) \\ \mathcal{L} f(\vb*{x}_2^{(\alpha)}) \\ \vdots \\ \mathcal{L} f(\vb*{x}_M^{(\alpha)})].
\end{equation}
Note that for collocated grids, we have $\alpha=\beta$. 
For this arrangement, the $j$th row of the matrix $\vb*{D}_{\mathcal{L}}^{(\alpha,\beta)}$ approximates the derivative at node $\vb*{x}^{(\alpha)}_j$ using the $n\ll N$ nearest nodes of the $\beta$-grid as the stencil for equation (\ref{rbf_poly_compact}).
The remaining weights are set to zero. 
As argued in \cite{flyer2016role}, the use of local RBF-FD approximations 
has the additional advantage that each stencil has its own supporting polynomial that can be locally adjusted.
The fully assembled matrix, $\vb*{D}_{\mathcal{L}}^{(\alpha,\beta)} $, is sparse with $M\times n$ nonzero elements. 
In this work, we demonstrate the use of these RBF-based differentiation matrices for incompressible flow simulations with staggered nodes. The following section discusses the discretization of the computational domain and the numerical scheme.

\section{Governing equations and numerical approach} \label{Spatial Discretization}

The motion of a general incompressible two-dimensional Newtonian fluid is governed by the Navier-Stokes and continuity equations,
\begin{align}
    &\pdv{u}{t} = -\left(u\pdv{u}{x}+v\pdv{u}{y}\right) -\pdv{p}{x}+\frac{1}{\mathrm{Re}}\left(\pdv{^2 u}{x^2}+\pdv{^2 u}{y^2}\right),\label{x_mom}\\
        &\pdv{v}{t} = -\left(u\pdv{v}{x}+v\pdv{v}{y}\right) -\pdv{p}{y}+\frac{1}{\mathrm{Re}}\left(\pdv{^2 v}{x^2}+\pdv{^2 v}{y^2}\right),\label{y_mom}\\
        &\pdv{u}{x}+\pdv{v}{y}=0.
\end{align}
All variables are nondimensionalized by the velocity scale $V$ and the length scale $L$, and $\mathrm{Re}$ denotes the Reynolds number.
Numerical instabilities are a known problem of standard FD methods that uses the Cartesian grid. This error can be traced back to central differencing schemes on collocated grids \cite{patankar2018numerical}. 
Similar grid oscillations are also observed on unstructured meshes. The arguably most common strategy to address this issue is the use of hyperviscosity \citep{flyer2012guide,fornberg2015solving,fornberg2011stabilization,shankar2018hyperviscosity}.
Instead of \emph{ad-hoc} regularization, we propose the use of a staggered node arrangement that discretizes the pressure in the computational domain $\Omega$ by $M$ scattered nodes that form the $P-$grid, and the velocity by $N$ scattered nodes that form the $V-$grid.
The $P-$grid is obtained using the Matlab algorithm {\tt DistMesh} developed by \citep{persson2005mesh}. This algorithm generates unstructured triangular meshes in 2-D. Inspired by standard FV methods that define the flux across the cell boundaries, we arrange the velocity components at the midpoints of cell edges. The resulting $N$ scattered nodes are the $V-$grid.
This staggered node arrangement is different from the classical staggered grid, which evaluates the horizontal and vertical velocities at different locations, but is similar to that used in FV methods.
The resulting ratio $N/M$ is around 3.

\begin{figure}[hbt!]
\centering
\includegraphics[trim = 0mm 2mm 0mm 8mm, clip, width=0.7\textwidth]{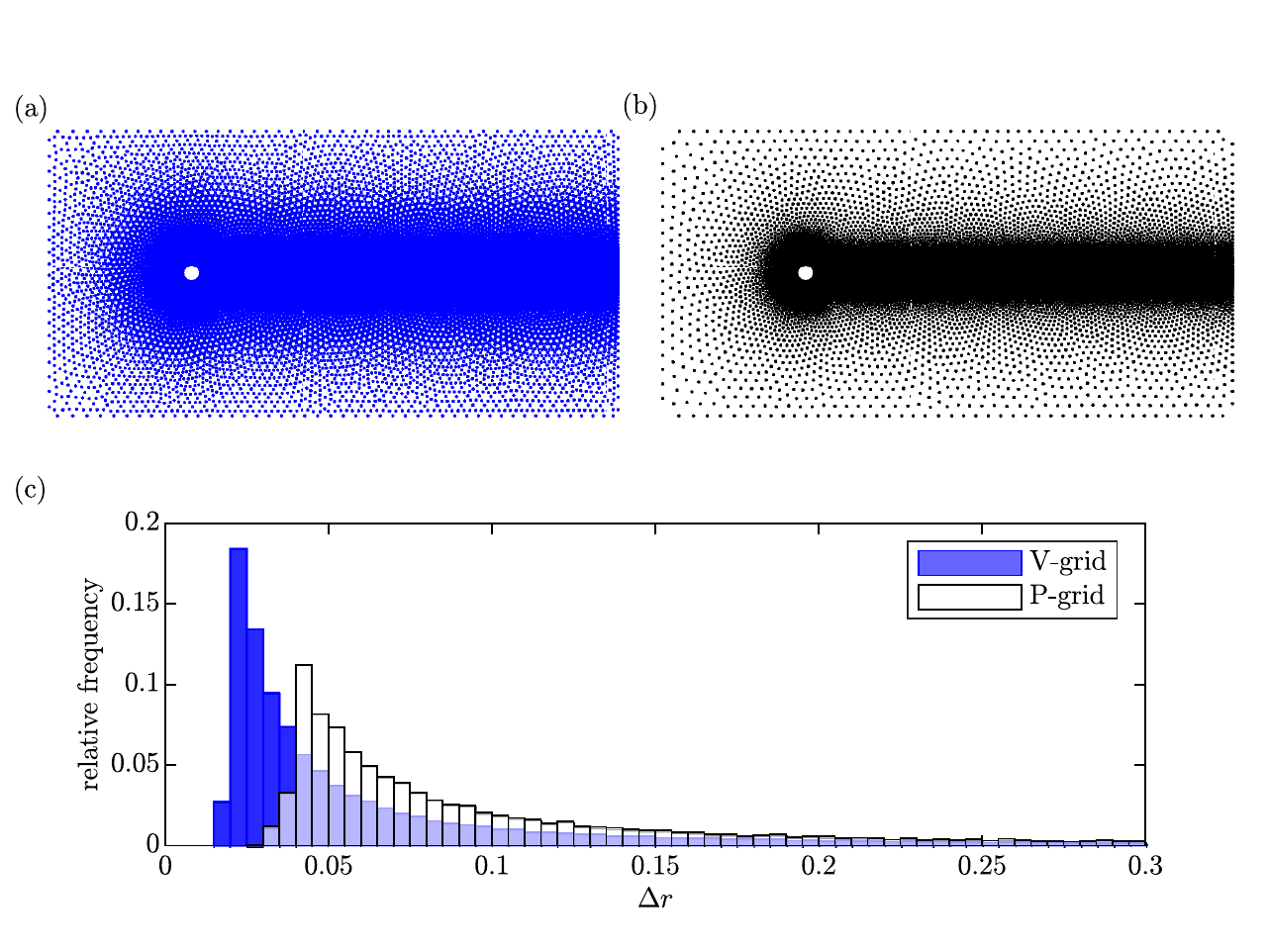}
\caption{Unstructured staggered node layout for flow around a cylinder: (a) V-grid; (b) P-grid; (c) grid spacing histograms. This grid is used for both the accuracy and error analyses in \S \ref{error_analysis}, and the simulations of the cylinder wake in \S \ref{cylinder flow}.}\label{staggered}
\end{figure}

Panels \ref{staggered}(a,b) shows this unstructured staggered node arrangement for the example of a cylinder in a rectangular domain with increased resolution close to the cylinder and in the wake region.
For each set of nodes that form a local stencil, a characteristic length $\Delta r$ can be determined as the locally averaged distance between adjacent nodes. The histograms of $\Delta r$ for the $V-$ and $P-$grids, shown in figure \ref{staggered}(c), indicate that the two grids are highly heterogeneous. The smallest characteristic spacing used to resolve the boundary layer near the cylinder differs by about one order of magnitude from the largest spacing in the far-field.

Several variants of the original fractional-step method by \cite{kim1985application}
using RBF-FD discretizations on collocated Cartesian grids \cite{javed2014shape,xie2021improved} and unstructured grids \cite{shahane2021high,unnikrishnan2021shear} can be found in literature. 
These previous implementations do not require regularizations, but, as we will demonstrate, significant savings in terms of the total number of grid points and stencil size is achieved by using the staggered node layout. We propagate the flow field from $j-$th time step to $(j+1)-$th time step by the following three-stage semi-implicit approach. The primitive variables $u,v,p$ are expressed in vector form as $\vb*{u},\vb*{v},\vb*{p}$, respectively. 
\begin{enumerate}
    \item 
    In the first stage, we use the second-order Adam-Bashforth method to discretize equations (\ref{x_mom}) and (\ref{y_mom}) in time explicitly, yielding
        \begin{align}
        \frac{\vb*{u}^*-\vb*{u}_{j}}{\Delta t} = \frac{3}{2}\vb*{C}_j(\vb*{u}_j)-\frac{1}{2}\vb*{C}_{j-1}(\vb*{u}_{j-1}), \qquad  \frac{\vb*{v}^*-\vb*{v}_{j}}{\Delta t} = \frac{3}{2}\vb*{C}_j(\vb*{v}_j)-\frac{1}{2}\vb*{C}_{j-1}(\vb*{v}_{j-1}).
    \end{align}
Here the convective term, $\vb*{C}_j$, is defined as  
    \begin{align}
               \vb*{C}_j(\vb*{q})&= -\left[ \vb*{u}_j\circ\left(\vb*{D}^{(V,V)}_x\vb*{q}\right) +\vb*{v}_j\circ \left( \vb*{D}^{(V,V)}_y \vb*{q}\right)\right],
    \end{align}
 where $\vb*{q}\in \{\vb*{u}, \vb*{v}\}$, superscripts $(\cdot)^*$ and $(\cdot)^{**}$ denote intermediate, non-divergence-free velocity fields, and $\circ$ the Hadamard product.
    \item 
    In the second stage, the viscous terms are advanced by the second-order implicit Crank-Nicolson scheme in time as
    \begin{align}\label{u_ss}
        \left(\vb*{I}-\frac{\Delta t}{2 \mathrm{Re}}\vb*{D}^{(V,V)}_{\Delta}\right)\vb*{u}^{**}=  \left(\vb*{I}+\frac{\Delta t}{2 \mathrm{Re}}\vb*{D}^{(V,V)}_{\Delta}\right)\vb*{u}^{*}, \qquad  \left(\vb*{I}-\frac{\Delta t}{2 \mathrm{Re}}\vb*{D}^{(V,V)}_{\Delta}\right)\vb*{v}^{**}=  \left(\vb*{I}+\frac{\Delta t}{2 \mathrm{Re}}\vb*{D}^{(V,V)}_{\Delta}\right)\vb*{v}^{*},
    \end{align}
    where $\vb*{I}$ denotes the identity matrix.
  The use of the Crank-Nicolson scheme eliminates the diffusive time-step constraint while maintaining second-order accuracy, see \ref{timestep} for more details.
    \item 
    In the third stage, incompressibility is enforced via pressure correction. First, we calculate the divergence of the intermediate velocity, $(\cdot)^{**}$, on the $P$-grid as
    \begin{align} \label{divergence}
        \vb*{F}_{j+1} = \vb*{D}^{(P,V)}_x \vb*{u}^{**} + \vb*{D}^{(P,V)}_y \vb*{v}^{**}.
    \end{align}
   Upon solution of the pressure-Poisson equation,
    \begin{align}\label{pressure_Poisson}
       \vb*{D}^{(P,P)}_{\Delta} \tilde{\vb*{p}}_{j+1} = \frac{1}{\Delta t} \vb*{F}_{j+1},
    \end{align}
    for the pressure correction, $\tilde{\vb*{p}}$, the pressure is obtained as
    \begin{align}
         \vb*{p}_{j+1} = \tilde{\vb*{p}}_{j+1} -\frac{\Delta t}{2\mathrm{Re}}  \vb*{D}^{(P,P)}_{\Delta} \tilde{\vb*{p}}_{j+1}.
         \end{align}
    The velocity components at the $(j+1)$-th time step are then calculated as
    \begin{align} \label{velocity_next_timestep}
         \vb*{u}_{j+1}  =\vb*{u}^{**} -\Delta t\vb*{D}^{(V,P)}_x \tilde{\vb*{p}}_{j+1}, \qquad  \vb*{v}_{j+1}= \vb*{v}^{**}-\Delta t\vb*{D}^{(V,P)}_y \tilde{\vb*{p}}_{j+1}.
    \end{align}
\end{enumerate}
Following \citet{kim1985application}, the boundary conditions for the intermediate velocities are
    \begin{align}
         \vb*{u}^{**}=\vb*{u}_{j+1} +\Delta t\vb*{D}^{(V,P)}_x \tilde{\vb*{p}}_{j} , \qquad  \vb*{v}^{**}= \vb*{v}_{j+1}+\Delta t\vb*{D}^{(V,P)}_y \tilde{\vb*{p}}_{j}.
    \end{align}
The matrix inversions in equations (\ref{u_ss}) and (\ref{pressure_Poisson}) are solved only once at the beginning using LU factorization 
with time complexity O($\frac{2}{3}N^3$) and O($\frac{2}{3}M^3$), respectively. At each time step, the time complexity is O($\max \{N^2,M^2\}$).

\section{Accuracy and error analysis} \label{error_analysis}

The spatial discretization scheme introduced in \S 3 above requires four types of differentiation matrices:
\begin{enumerate}
    \item $\vb*{D}^{(V,V)}_x$, $\vb*{D}^{(V,V)}_y$, and $\vb*{D}^{(V,V)}_{\Delta}$ with dimension $N\times N$ from V-grid to V-grid;
    \item $\vb*{D}^{(P,V)}_x$, $\vb*{D}^{(P,V)}_y$ with dimension $M\times N$ from V-grid to P-grid;
    \item $\vb*{D}^{(P,P)}_{\Delta}$ with dimension $M\times M$ from P-grid to P-grid;
    \item $\vb*{D}^{(V,P)}_x$, $\vb*{D}^{(V,P)}_y$ with dimension $N\times M$ from P-grid to V-grid.
\end{enumerate}

\begin{figure}[hbt!]
\centering
\includegraphics[trim = 10mm 15mm 15mm 0mm, clip,width=0.7\textwidth]{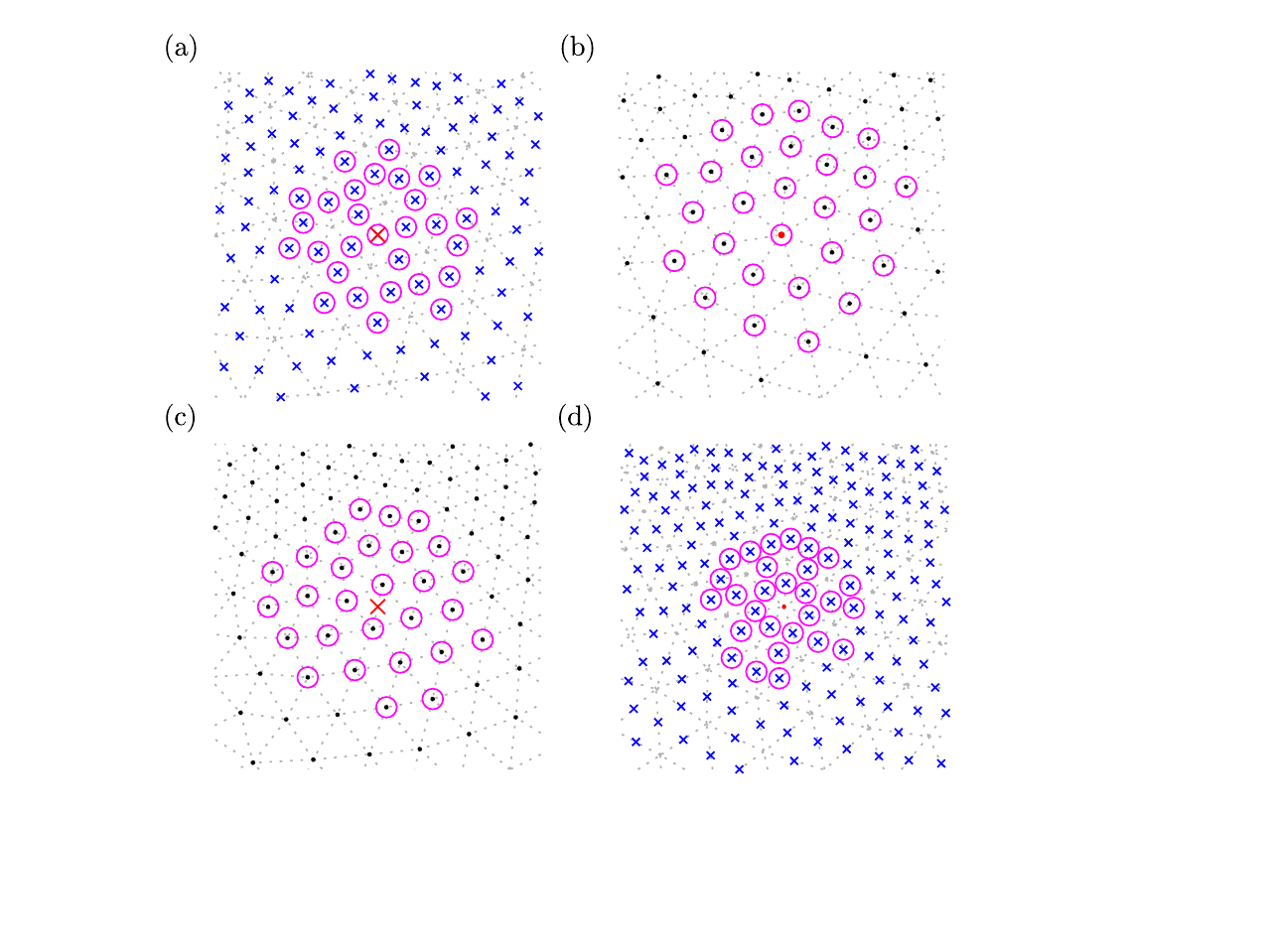}
\caption{Stencils (magenta circles) for differentiation matrices: (a) $\vb*{D}^{(V,V)} $ ; (b) $\vb*{D}^{(P,P)} $; (c) $\vb*{D}^{(V,P)} $ ; (d) $\vb*{D}^{(P,V)}$. A constant stencil size $n=28$ is used to approximate the derivatives at given nodes (red). Referred to figure \ref{staggered} for the markers. The underlying triangular mesh is shown as gray dotted lines.}\label{mesh_neighbor}
\end{figure}

Refer to figure \ref{staggered} for the grid topology. For any given node, the RBF stencil consists of the nearest $n$ nodes. For $n=28$, figure \ref{mesh_neighbor} shows examples of stencils used for the four different cases. In panels \ref{mesh_neighbor}(a) and (b), $\alpha=\beta$ and the node of interests hence is a node on the grid. In panels \ref{mesh_neighbor}(c) and (d), on the other hand, the nodes
comprising the stencil are from different grids.

In the following, we propose an error analysis strategy that minimizes the truncation error of the spatial derivatives on the staggered grid
by selecting an appropriate combination of the stencil size, $n$, the exponent of the PHS, $m$, and the degree of polynomials, $q$.
In previous studies, the discretization error for the augmented RBF-PHS method has been examined in terms of the convergence rate under grid refinement \cite{bayona2017role,flyer2016enhancing,flyer2016role,shahane2021high}.
A general observation is that the error convergence depends on the degree of the polynomial, and a minimum stencil size is required for numerical stability.
In this work, we propose the use of modified wavenumber analysis, which we generalize to scattered nodes.
Classical modified wavenumber analysis for lattice-based node sets is commonly used to measure the accuracy of finite difference schemes, see e.g., \citep{moin2010fundamentals}.
For a given lattice with grid spacing $\Delta x$, the modified wavenumber, $k^*$, is computed by applying the finite difference to the discretized
sinusoidal function $g(x)=\mathrm{e}^{\mathrm{i} k x}$. 
The difference between $k^*\Delta x$ and $k\Delta x$ indicates the numerical error as a function of wavenumber.
To be able to investigate the discretization error as a function of the wave angle, $\theta=\arctan{(k_y/k_x)}$, relative to the fixed set of scattered nodes, we define the transformed radial coordinate $\tilde{r}= (k_x x+k_y y)/\tilde{k}$, where $\tilde{k} = \sqrt{k_x^2+k_y^2}$ is the radial wavenumber. As in \cite{sengupta2011analysis, tan2021two}, the radial modified wavenumber is then given as
\begin{align} \label{2d_mw}
     k^* =   -\mathrm{i}   \delta_{\tilde{r}}\qty{\mathrm{e}^{\mathrm{i} \tilde{k}\tilde{r}}}   \mathrm{e}^{-\mathrm{i} \tilde{k}\tilde{r}},
\end{align}
where ${\delta}_{\tilde{r}}$ is the discrete RBF-FD differentiation operator. Equation (\ref{2d_mw}) reduces to the standard one-dimensional case for $k_x=0$ or $k_y=0$. Prior to performing the modified wavenumber analysis, we first identify the optimal parameters using a fixed nondimensional wavenumber, $k\Delta r= \sqrt{2}$, highlighted in figures \ref{mw_V2V} and \ref{mw_P2P}. This specific value is chosen to guarantee the `spectral-like resolution' \citep{lele1992compact} that is achieved by higher-order Pad\'e-type compact finite differences, and that has made the latter a popular choice for direct numerical simulation (DNS) \citep{colonius1997sound,joslin1993spatial,kloker1997robust,lee1993direct,samtaney2001direct,suzuki2003shock}.

The test function takes the form 
\begin{align}\label{test}
    g_j(\vb*{x}) =\cos{\left(\frac{x}{\Delta r_j}\right)}\cos{\left(\frac{y}{\Delta r_j}\right)},
\end{align}
where $j$ is the node index. The corresponding wavelength is $\lambda = \sqrt{2}\pi \Delta r_j$. The average relative error of the augmented RBF-PHS method is determined as 
\begin{align}\label{relative_error}
    E^{(\alpha,\beta)}  = \frac{1}{N} \sum_{j=1}^N \frac{\lvert {\delta}_j^{(\alpha,\beta)} g_j(\vb*{x}^{(\beta)})-\mathcal{L}g_j(\vb*{x}^{(\alpha)})\lvert}{\max\{\lvert \mathcal{L}g_j(\vb*{x})\lvert\}}, 
\end{align}
where $\delta_j$ represents the local RBF differentiation operation at the $j$th node.  
The local minima of the error guides the selection of parameters: 
$n$, the stencil size, $m$, the exponent of the PHS, and $q$, the polynomial degree. It has been shown by \cite{bayona2017role,flyer2016enhancing,flyer2016role,shahane2021high} that $q$ determines the overall order of accuracy. 

In the remainder of this section, we use as the test mesh for the error analysis the {\tt distmesh} grid shown in figure \ref{staggered} above. Due to its highly non-uniform node distribution and large variation of $\Delta r$, $0.03 <\Delta r< 0.79$, it is representative of meshes used to discretize complex geometries. This ensures that the error estimates are conservative and applicable to non-generic scenarios.

Our requirements for the Navier-Stokes solver are: 
\begin{enumerate}[(i)]
    \item compact stencil size of $n\lesssim 30$ for numerical efficiency,
    \item formal order-of-accuracy of $q\geq 2$ for physical accuracy,
    \item small relative error,
    \item stability (requires $n\gtrsim(q+1)(q+2)$ for scattered nodes, see e.g. \citep{bayona2019insight,bayona2017role,bayona2019role, flyer2016enhancing}).
\end{enumerate}
To meet these criteria, we vary the different parameters and conduct detailed error analyses for the $(V,V)$- and $(P,P)$-grids, reported in \S \ref{V2V} and \S \ref{P2P}, respectively. The analyses for the remaining combinations are reported in \ref{V2P_P2V}, 
and a comparison with the established RBF-QR implementation by \citet{fornberg2011stabilization} and the RBF-GA implementation by \citet{bollig2012solution} is shown in \ref{RBF_comparison}.

\subsection{V-grid to V-grid} \label{V2V}

\begin{figure}[hbt!]
\centering
\includegraphics[trim = 5mm 8mm 5mm 5mm, clip, width=0.75\textwidth]{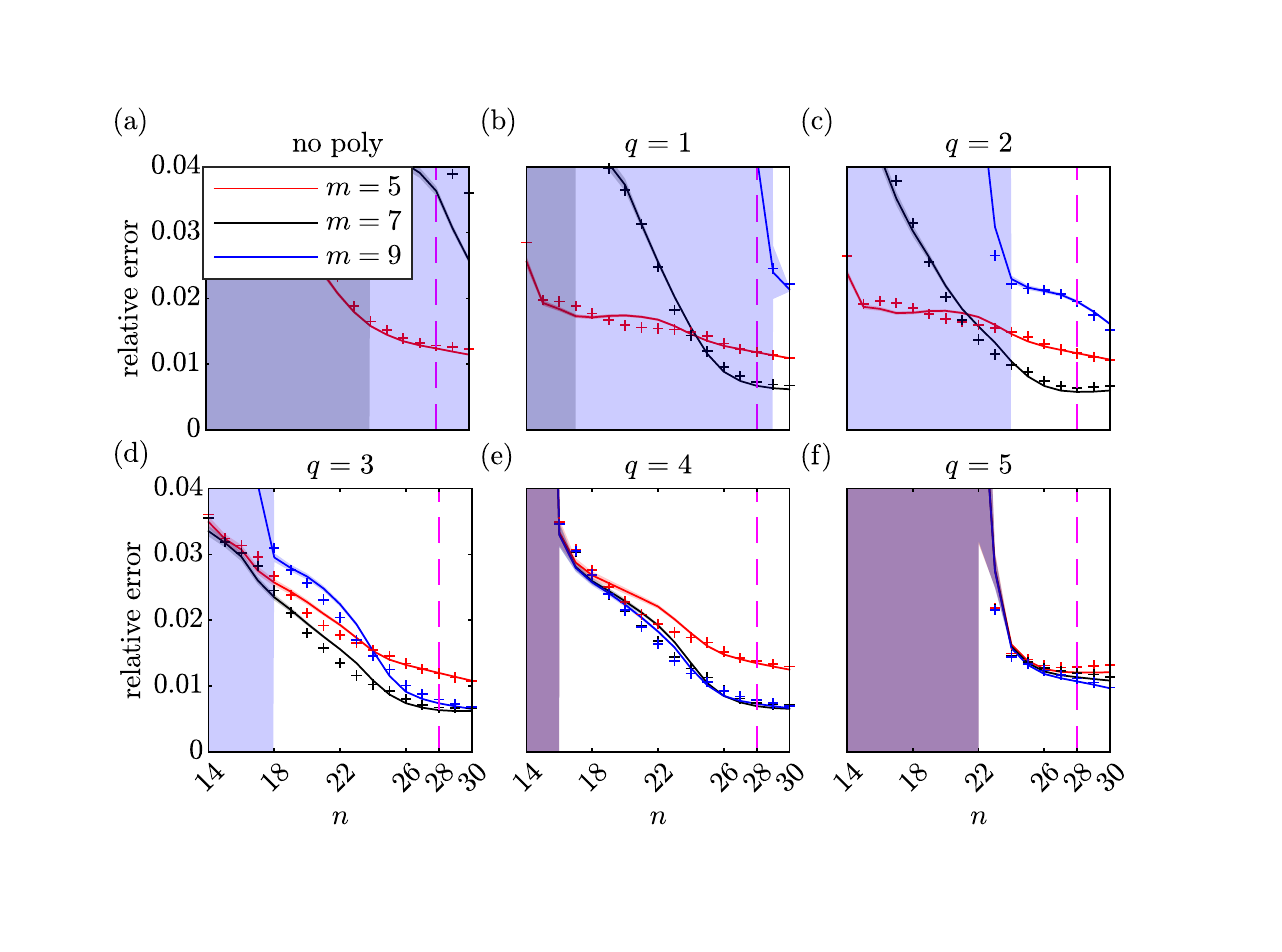}
\caption{Relative error for $\vb*{D}^{(V,V)}_x$ 
for different combinations of PHS exponents, $m$, and  polynomial orders, $q$: (a) no polynomial augmentation; (b) $q=1$; (c) $q=2$; (d) $q=3$; (e) $q=4$; (f) $q=5$. The results for the $y$-derivatives (`+') are almost indistinguishable. Shaded areas of the same color show the standard deviation (overlapping regions appear purple).}\label{error_dx_V2V}
\end{figure}

We first consider the differentiation matrices $\vb*{D}_{x}^{(V,V)}$ and $\vb*{D}_{\Delta}^{(V,V)}$, which operate on the $(V,V)$-grid shown in figure \ref{mesh_neighbor}(a).
Figure \ref{error_dx_V2V} shows the average relative errors, equation (\ref{relative_error}), computed for the first derivative in the $x$-direction for the test function, equation (\ref{test}). 
By comparing the different subplots of figure \ref{error_dx_V2V}, we observe that the relative error for $m=5$ is largely independent of the polynomial order, $q$. For $q\leq 3$, the relative error for $m=5$ is larger than for $m=7$ and $9$, and we hence do not further consider $m=5$.
A general trend is that the truncation error decreases with increasing stencil size and then stagnates. 
Similar results are also found in \citet{bayona2017role}.
The last observation is that for $q=5$, the relative error has increased for all $m$. To avoid the Runge phenomenon near the domain boundary, a smaller value of $q$ is favored for a fixed stencil size \citep{bayona2019insight,bayona2017role,bayona2019role}.
With the goal to minimize both the stencil size error, we identify $n=28$, $m=7$, and $q=3$ as the best combination. Note that the order of accuracy is not reflected in this analysis of the relative error for a fixed $k\Delta r$.
The modified wavenumber analysis (shown in figures \ref{mw_V2V} and \ref{mw_q4} below) will, however, show that increasing $q$ beyond $3$ does not improve the order of accuracy for a fixed stencil size, $n=28$.


\begin{figure}[hbt!]
\centering
\includegraphics[trim = 5mm 12mm 5mm 5mm, clip, width=0.75\textwidth]{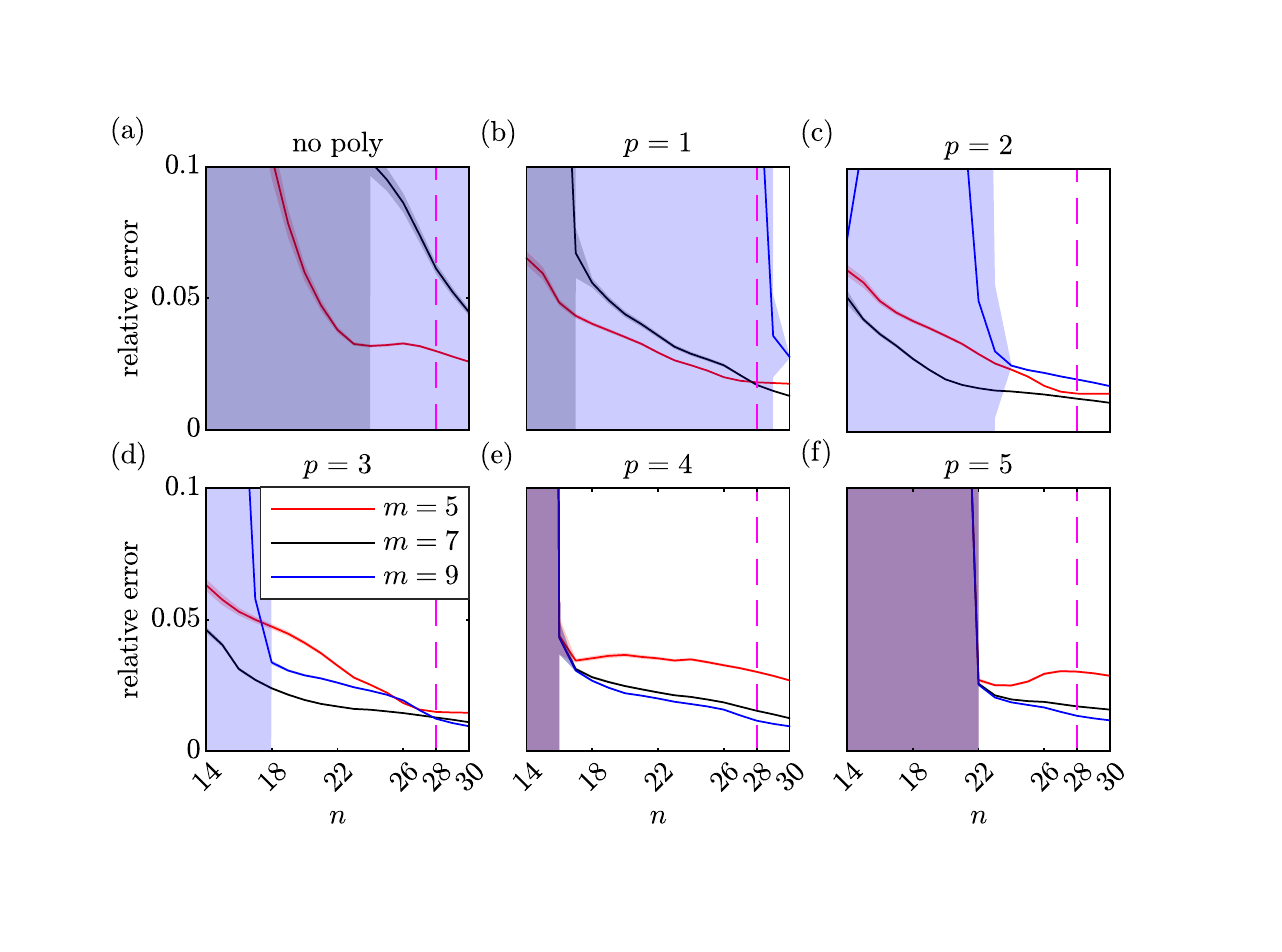}
\caption{Relative error for $\vb*{D}^{(V,V)}_{\Delta}$ for different combinations of 
PHS exponents, $m$, and  polynomial orders, $q$: (a) no polynomial; (b) $q=1$; (c) $q=2$; (d) $q=3$; (e) $q=4$; (f) $q=5$. Shaded areas of the same color show the standard deviation (overlapping regions appear purple).}\label{error_L_V2V}
\end{figure}

Figure \ref{error_L_V2V} shows the relative error for the second derivatives contained in the Laplacian, $\vb*{D}^{(V,V)}_{\Delta}$. 
We repeat the same analysis as for the first derivative, and similar trends are observed.
Following the same arguments stated above, and for consistency, we proceed with $(n,m,q) = (28,7,3)$ for the Laplacian as well.

\begin{figure}[hbt!]
\centering
\includegraphics[trim = 10mm 20mm 10mm 15mm, clip, width=0.7\textwidth]{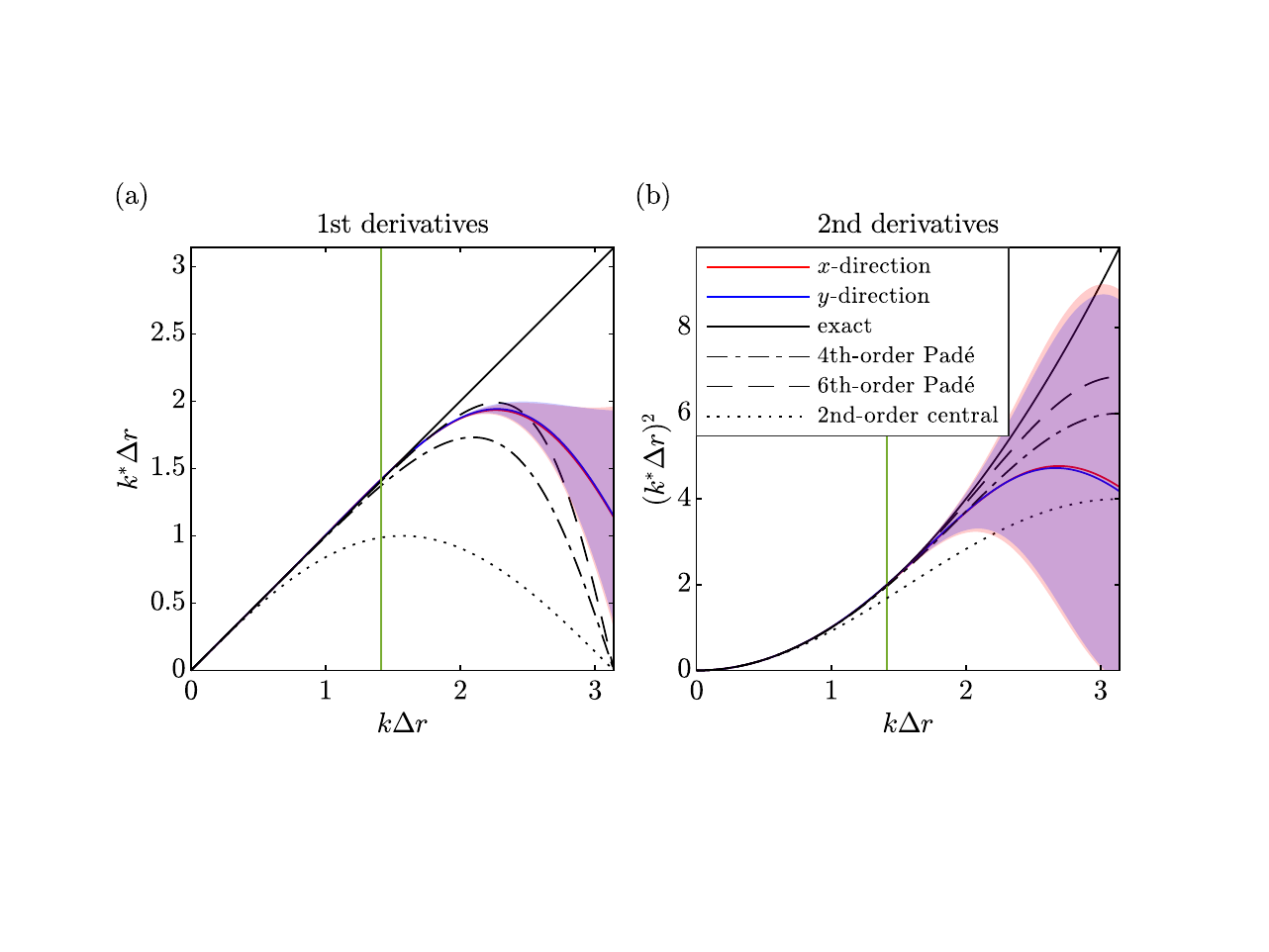}
\caption{Modified wavenumber diagrams for the differentiation matrices $\vb*{D}^{(V,V)}$ with $(n,m,q)=(28,7,3)$ for: (a) the first; (b) the second derivatives. Shaded areas represent the standard deviations for the results in the $x$- (red) and $y$- (blue) directions, respectively (overlapping regions appear purple).
Results for Pad\'e-type methods are shown for comparison. The recommended maximum modified wavenumber of $k\Delta r=\sqrt{2}$ is highlighted in green. }\label{mw_V2V}
\end{figure}

Figure \ref{mw_V2V} shows the modified wavenumber diagrams for the differentiation matrices $\vb*{D}^{(V,V)}$ with the corresponding standard deviations. Results for the 
standard $2$nd-order central, and $4$th- and $6$th-order Pad\'e-type methods are shown for comparison.
Despite the inhomogeneity of the mesh, the truncation errors in the $x$ and $y$ directions are very similar.
The accuracy of the first and second derivatives is almost identical to the $6$th-order Pad\'e scheme up to $k\Delta r=2$. Also shown is the standard deviation of the modified wavenumbers obtained from all grid points.
The important observation is that the standard deviation is negligible until the mean modified wavenumber deviates from the theoretical curve. 
Wavenumbers beyond this point have to be considered underresolved and have to be filtered out or, ideally, avoided entirely through grid refinement. 
The proposed maximum modified wavenumber of $k\Delta r=\sqrt{2}$ (green lines in figure \ref{mw_V2V}) ensures the spectral-like accuracy, which exceeds the second-order accuracy (dotted lines) of most commercial and general-purpose computational fluid mechanics codes.

\begin{figure}[hbt!]
\centering
\includegraphics[trim = 15mm 20mm 0mm 15mm, clip, width=0.7\textwidth]{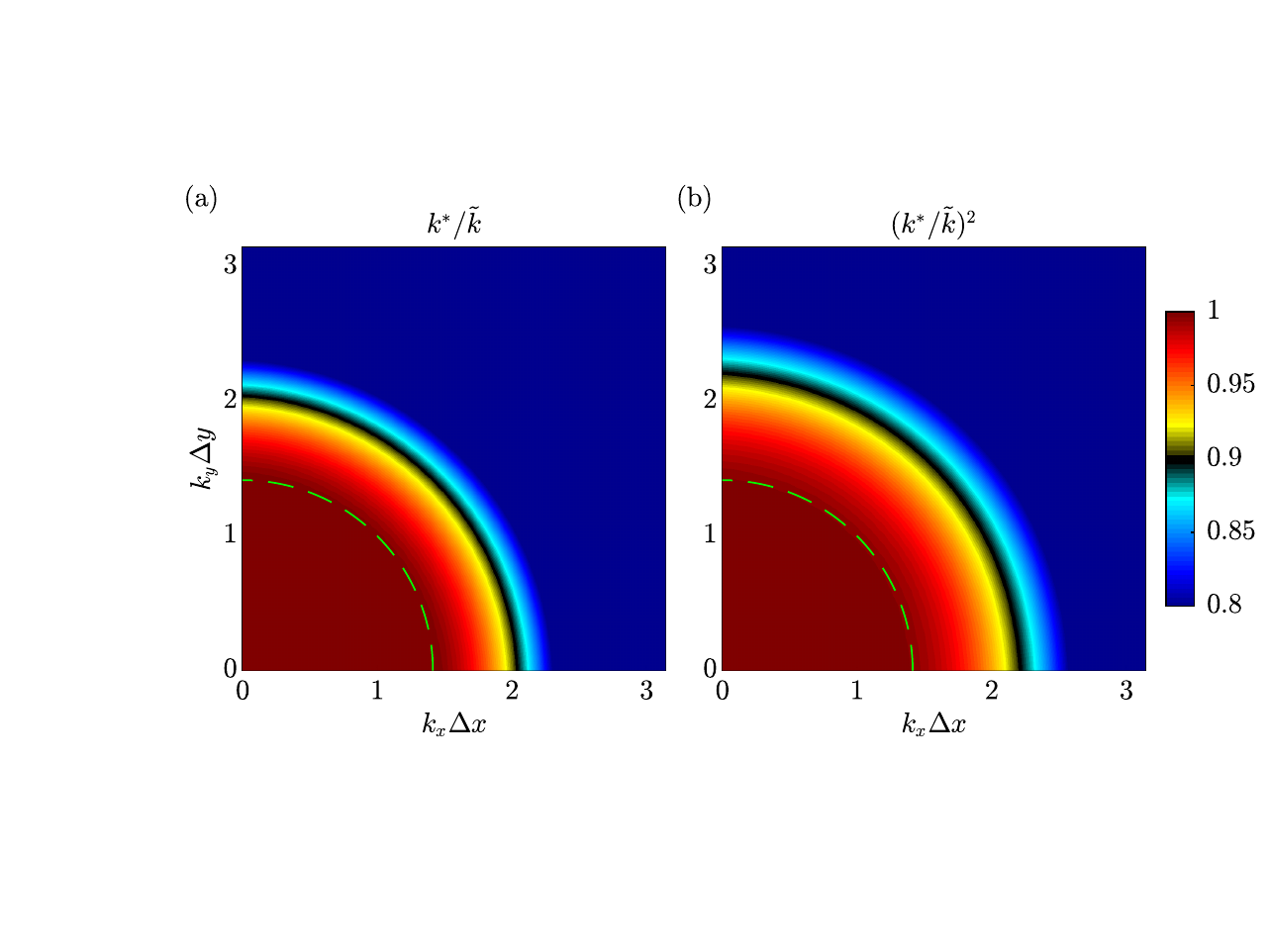}
\caption{2D modified wavenumber diagrams (normalized) for the differentiation matrices $\vb*{D}^{(V,V)} $ with $(n,m,q)=(28,7,3)$ : (a) first and (b) second derivatives. The recommended maximum modified wavenumber of $\tilde{k}\Delta r=\sqrt{2}$ is highlighted in green.  }\label{mw_V2V_2d}
\end{figure}

To assess the inhomogeneity of the computational grid, we extend the modified wavenumber analysis to two dimensions by plotting the ratio $k^*/\tilde{k}$, where $\tilde{k}=\sqrt{k_x^2+k_y^2}$, in the $x$-$y$ wavenumber plane. This is shown in figure \ref{mw_V2V_2d}.
An important observation is that the normalized modified wavenumbers are almost independent of the direction of the wave. This finding is encouraging since the local topology of an unstructured mesh, in principle, can have arbitrary orientations.
At the proposed maximum modified wavenumber of $\tilde{k}\Delta r=\sqrt{2}$ (green lines), 
 the ratio deviates from the spectral limit by at most
 $0.6\%$ and $0.2\%$  for the first and second derivatives, respectively. 
These observations suggest that the 28-point PHS+poly RBF-FD discretization provides high-order accuracy independent of grid orientation
if the nodes are taken as the midpoints of a {\tt{DistMesh}} grid, see figure \ref{staggered}.

\subsection{P-grid to P-grid} \label{P2P}

\begin{figure}[hbt!]
\centering
\includegraphics[trim = 5mm 12mm 5mm 5mm, clip, width=0.75\textwidth]{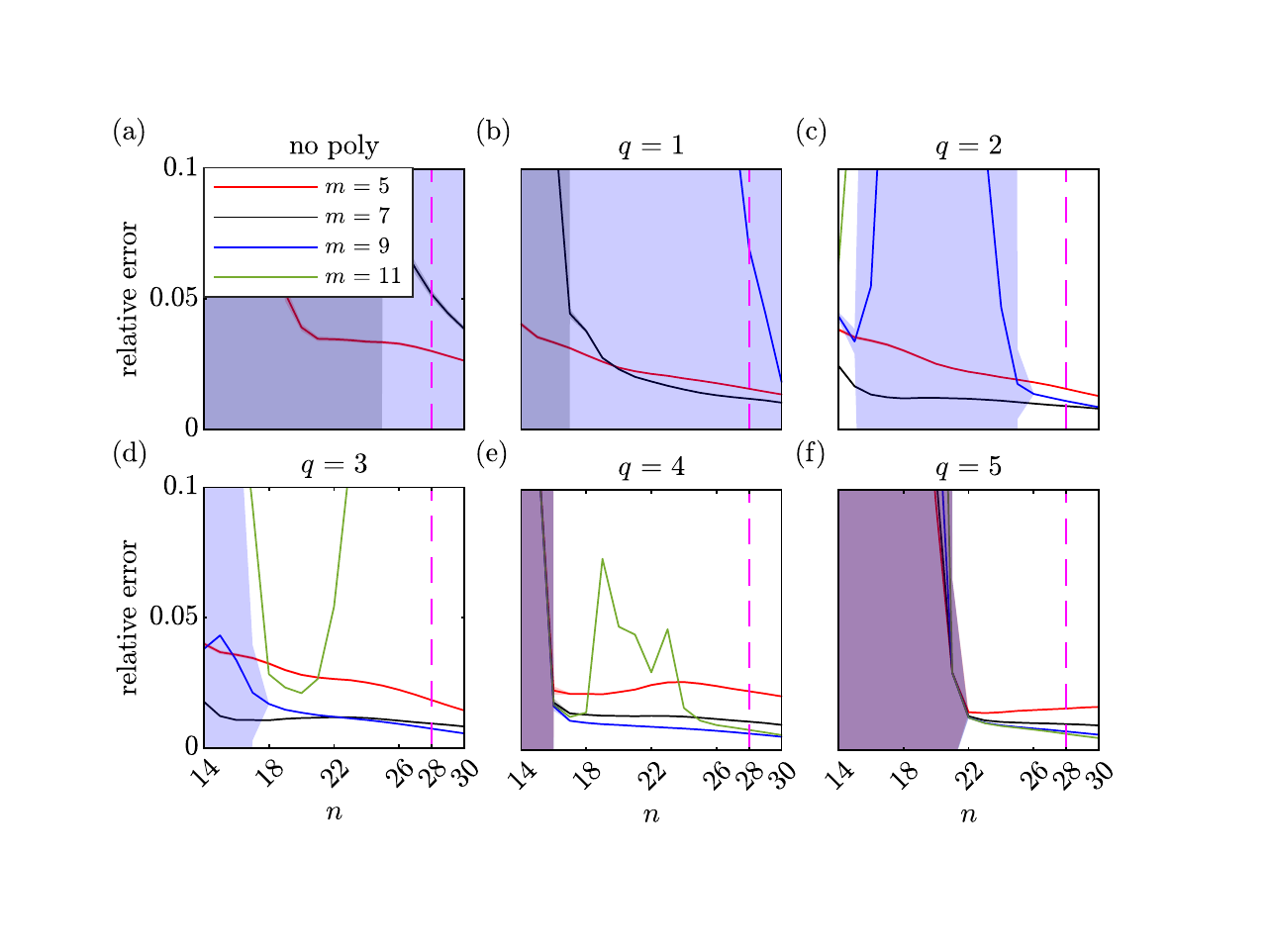}
\caption{Relative error for $\vb*{D}^{(P,P)}_{\Delta}$ for different combinations of PHS exponents, $m$, and  polynomial orders, $q$: (a) no polynomial; (b) $q=1$; (c) $q=2$; (d) $q=3$; (e) $q=4$; (f) $q=5$. Shaded areas of the same color show the standard deviation (overlapping regions appear purple).}\label{error_L_P2P}
\end{figure}

Next, we repeat the analysis of \S \ref{V2V} for the differentiation matrix $\vb*{D}_{\Delta}^{(P,P)}$, which operates on the $(P,P)$-grid shown in figure \ref{mesh_neighbor}(b).
We focus on the truncation error of the discretized Laplace operator that is used to solve the pressure Poisson problem, equation \ref{pressure_Poisson}.
Figure \ref{error_L_P2P} shows the relative error for different parameters. The combination $(n,m,q)=(28,7,3)$ that was identified as optimal for the $(V,V)$-grid is taken as the baseline.
Deviating from the baseline case, increasing $m$ to $m=9$ with $q$ constant or increasing $q$ to $q=4$ with $m$ constant yields a small decrease of the error. We choose consistency over these marginal gains and proceed with $(n,m,q)=(28,7,3)$ for $\vb*{D}_{\Delta}^{(P,P)}$.  
Prior to making this choice, we confirmed that further increasing $m$ does not further decrease the error significantly. 
In fact, letting $m=11$ leads to numerical instability, as can be seen in figure \ref{error_L_P2P}(d).


\begin{figure}[hbt!]
\centering
\includegraphics[trim = 0mm 2mm 0mm 2mm, clip, width=0.75\textwidth]{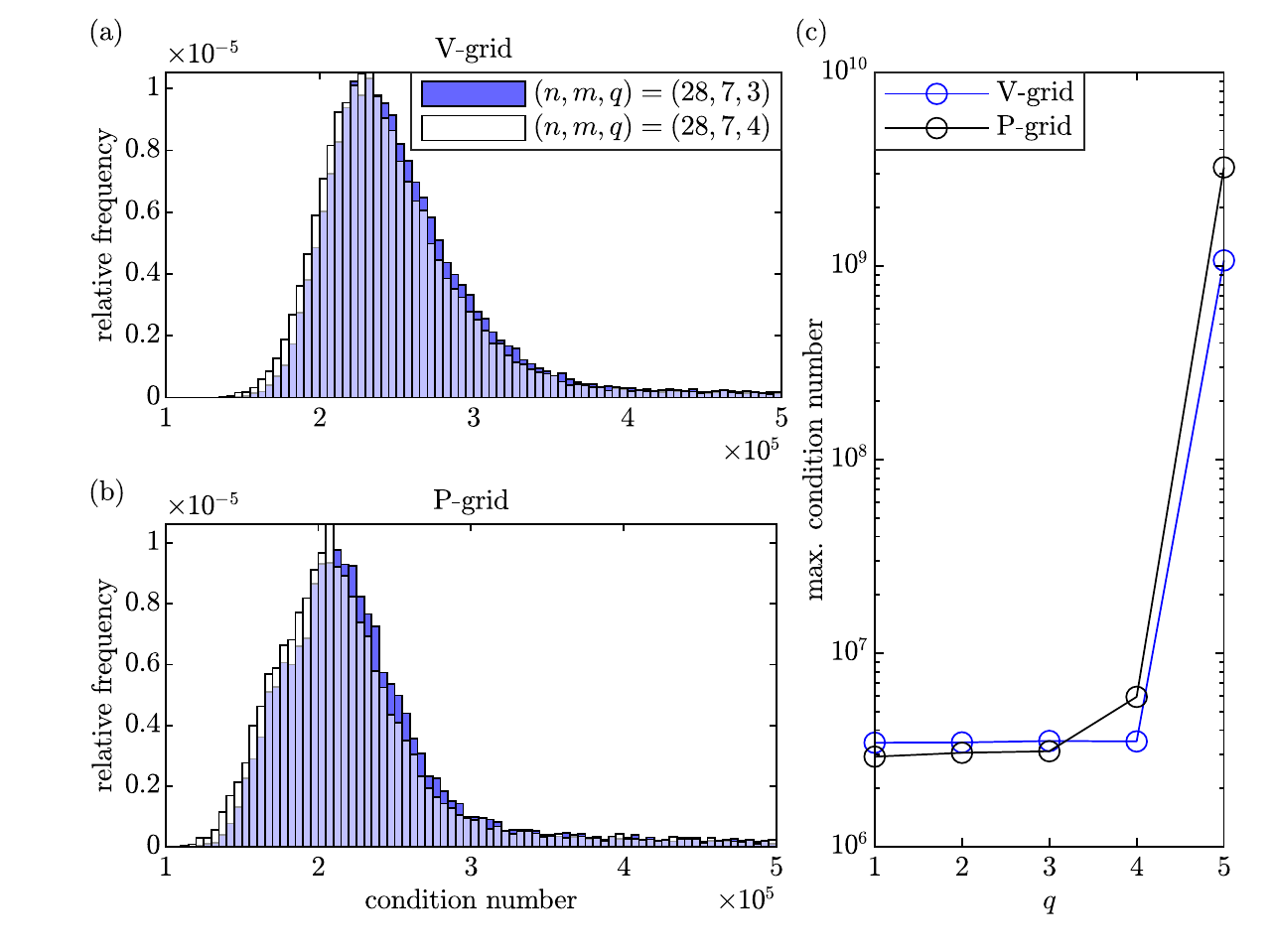}
\caption{Histogram of the condition numbers of $\vb*{A}_\text{aug}$: (a) V-grid and (b) P-grid; (c) maximum value for different polynomial degrees.
} \label{cond_number}
\end{figure}
To gauge the numerical stability of the local stencils, figure \ref{cond_number}(a,b) show the histograms of the condition number of $\vb*{A}_\text{aug}$, defined in equation (\ref{rbf_poly_compact}), for $q=3$ and $q=4$ on $V-$ and $P-$grids. While the histograms are comparable, the maximum condition number for $P-$grid shown in panel \ref{cond_number}(c) reveals that the condition number starts to deteriorate for $q\geq4$. This is in accordance with the findings of \citet{flyer2016enhancing} who found that lower polynomial degrees lead to more stable discretizations. We will show in the following that increasing $q$ beyond $3$ also does not increase the accuracy for a fixed stencil size.

\begin{figure}[hbt!]
\centering
\includegraphics[trim = 10mm 20mm 10mm 15mm, clip, width=0.7\textwidth]{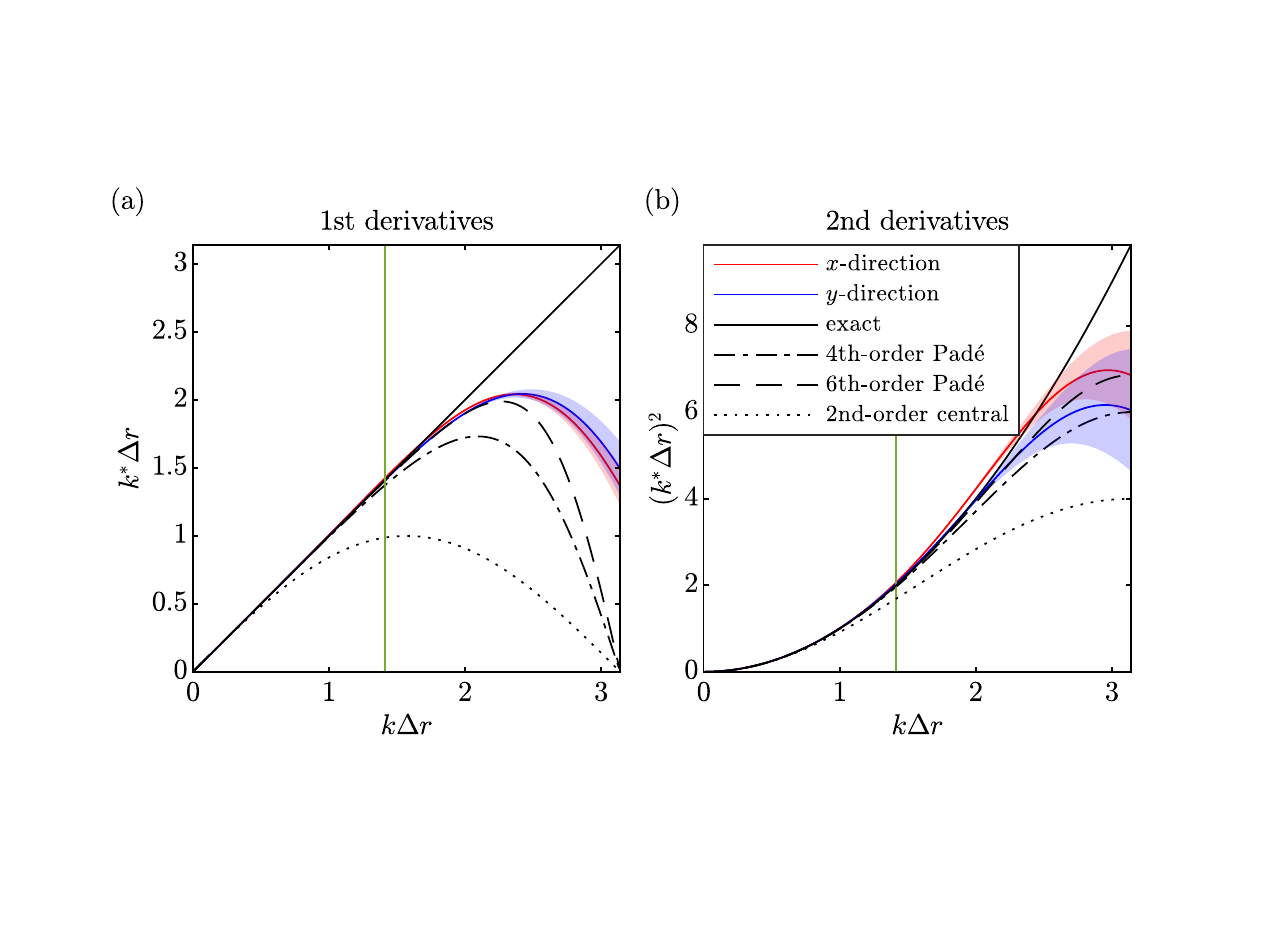}
\caption{
Same as figure \ref{mw_V2V} for $\vb*{D}^{(P,P)}$ with $(n,m,q)=(28,7,3)$. Results for $(n,m,q)=(28,7,4)$ are reported in panels \ref{mw_q4}(c,d) with negligible difference. 
}\label{mw_P2P}
\end{figure}

The modified wavenumber diagrams for the differentiation matrices $\vb*{D}^{(P,P)}$ are reported in figure \ref{mw_P2P}. As before, the analytical results for the $2$nd-order central, and $4$th- and $6$th-order Pad\'e-type methods are shown for comparison.
It is found that the modified wavenumber curves for the $(P,P)$-grid are as good or better as the $6$th-order Pad\'e scheme for both the first and second derivatives.
Furthermore, the variances are much lower compared to the results for the $(V,V)$-grid, previously shown in figure \ref{mw_V2V}. 
The likely reason is that the local topology of the $(P,P)$-grid is more evenly distributed in space. This becomes apparent when comparing figure \ref{mesh_neighbor}(b,c) to (a,d) of the same figure.
It is also observed that the modified wavenumber curves for 
the $x$- and $y$-directions, in particular for the second derivatives, differ. 
We suspect that this is due to the inhomogeneity of the computational domain, which results from its elongated shape and the local grid refinement of the wake region that is also oriented in the $x-$direction. 
We conclude that the discretization is also robust in the presence of grid non-homogeneity while preserving higher-order accuracy.

\begin{figure}[hbt!]
\centering
\includegraphics[trim = 15mm 20mm 0mm 15mm, clip, width=0.7\textwidth]{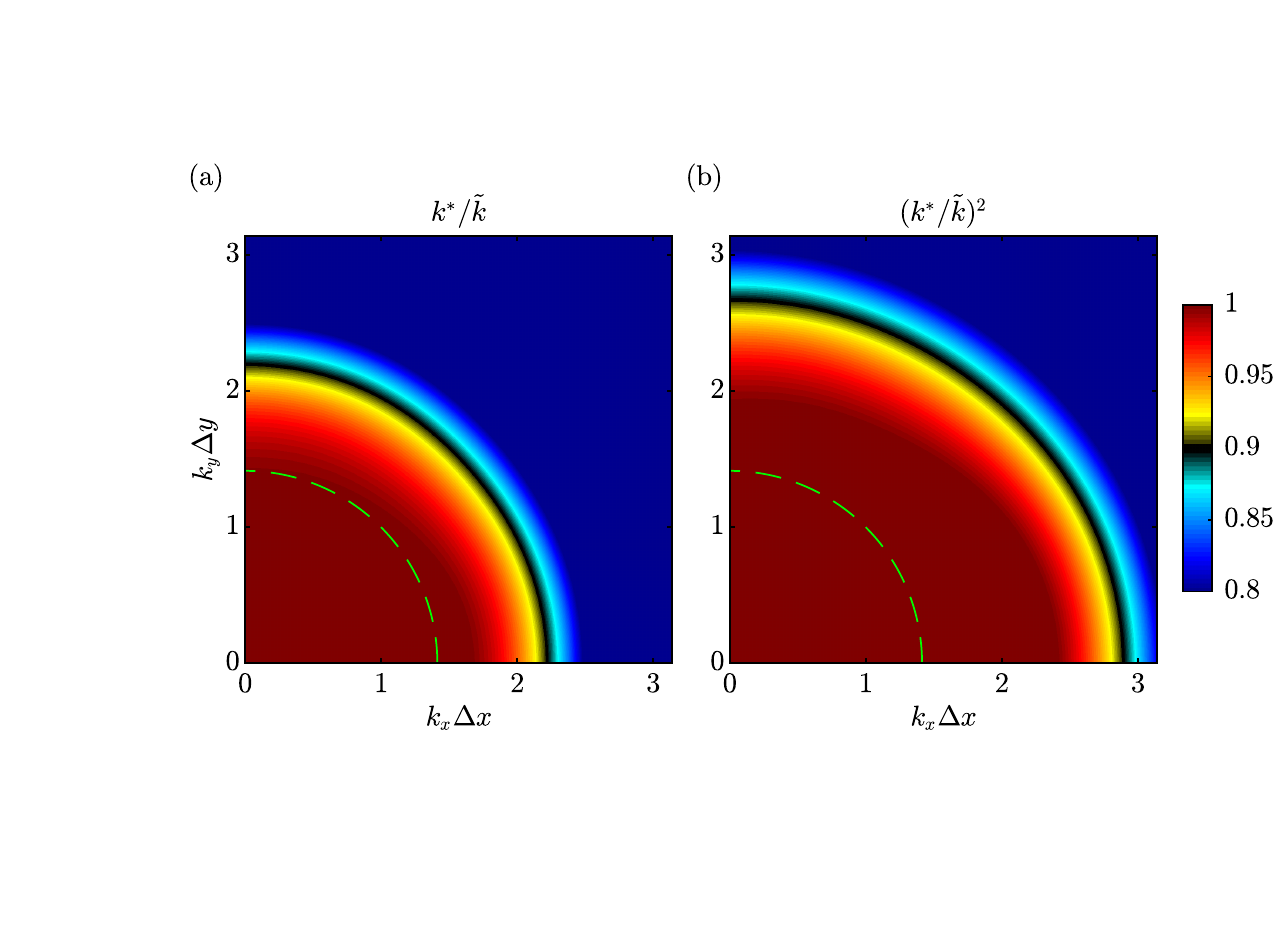}
\caption{Same as figure \ref{mw_V2V_2d} for $\vb*{D}^{(P,P)}$ with $(n,m,q)=(28,7,3)$.  }\label{mw_P2P_2d}
\end{figure}

The effect of grid orientation is assessed for the first and second derivatives in terms of the two-dimensional modified wavenumber diagrams in figure \ref{mw_P2P_2d}(a) and (b), respectively. Similar to what was found for the $(V,V)-$grid in figure \ref{mw_V2V_2d}, no strong directional preference is observed. The direct comparison with figure \ref{mw_V2V_2d} reveals that the modified wavenumber stays closer to unity of a larger radius. This implies that the $(P,P)-$grid will yield accurate results over a wider wavenumber range.

\begin{figure}[hbt!]
\centering
\includegraphics[trim = 0mm 2mm 0mm 42mm, clip, width=0.7\textwidth]{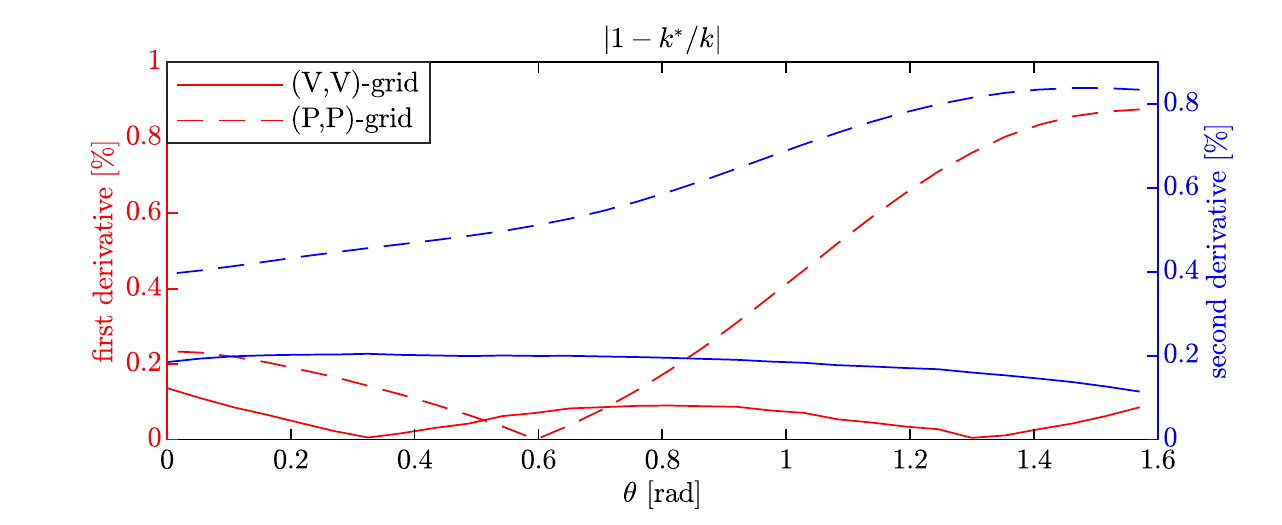}
\caption{
 Relative error of the modified wavenumber at different wave angles. }\label{MW_2d_error}
\end{figure}
Figure \ref{MW_2d_error} shows a direct comparison of the error relative to the spectral limit at $\tilde{k}\Delta r=\sqrt{2}$ as a function of the wave angle for both grids and derivatives. The errors for the $(P,P)$-grid is around $0.4\%$ and $0.6\%$ for first and second derivatives, respectively. The values for $(V,V)$-grid are generally lower. This shows that, despite the heterogeneous grid, the dependence on the wave angle is low.

The relative error analyses for $\vb*{D}^{(V,P)}$ and $\vb*{D}^{(P,V)}$, that is, cases (c) and (d) in figure \ref{mesh_neighbor}, are reported in \ref{V2P_P2V}. The results are similar to those of the non-mixed grids above. To maintain the overall accuracy of the numerical scheme, which involves combinations of all grids, we hence retain the recommended value of
$\tilde{k}\Delta r=\sqrt{2}$.

\section{Applications} \label{applications}

We demonstrate the viability, robustness, and flexibility of the fractional-step, staggered-node, PHS+poly RBF-FD algorithm on two examples, internal lid-driven cavity flow and the flow past a cylinder. Both test cases are established benchmark problems. The parameters identified from the error and accuracy analyses in the previous section, i.e., $(n,m,q)=(28,7,3)$, are used throughout, and it is ensured that the {\tt{DistMesh}} grid properly resolves the flow by adhering to the $\tilde{k}\Delta r\lesssim \sqrt{2}$ recommendation.

\subsection{Lid-driven cavity flows} \label{Lid-Driven cavity flows}

The lid-driven cavity problem is often used to validate
different implementations of RBF-FD methods. In the past, these implementations often used specific grid arrangements, such as Cartesian or quasi-uniform grids \citep{sanyasiraju2008local, xie2021improved}, or locally orthogonal grids near the boundary \citep{ chinchapatnam2009compact, shu2003local,shu2005computation}. A notable exception is the vorticity/stream function-based steady-state solver by \citet{bayona2017role}, which uses scattered nodes throughout. Here, we solve the transient problem on scattered nodes with local grid refinement near the walls.

\begin{figure}[hbt!]
\centering
\includegraphics[trim = 0mm 22mm 0mm 18mm, clip, width=0.85\textwidth]{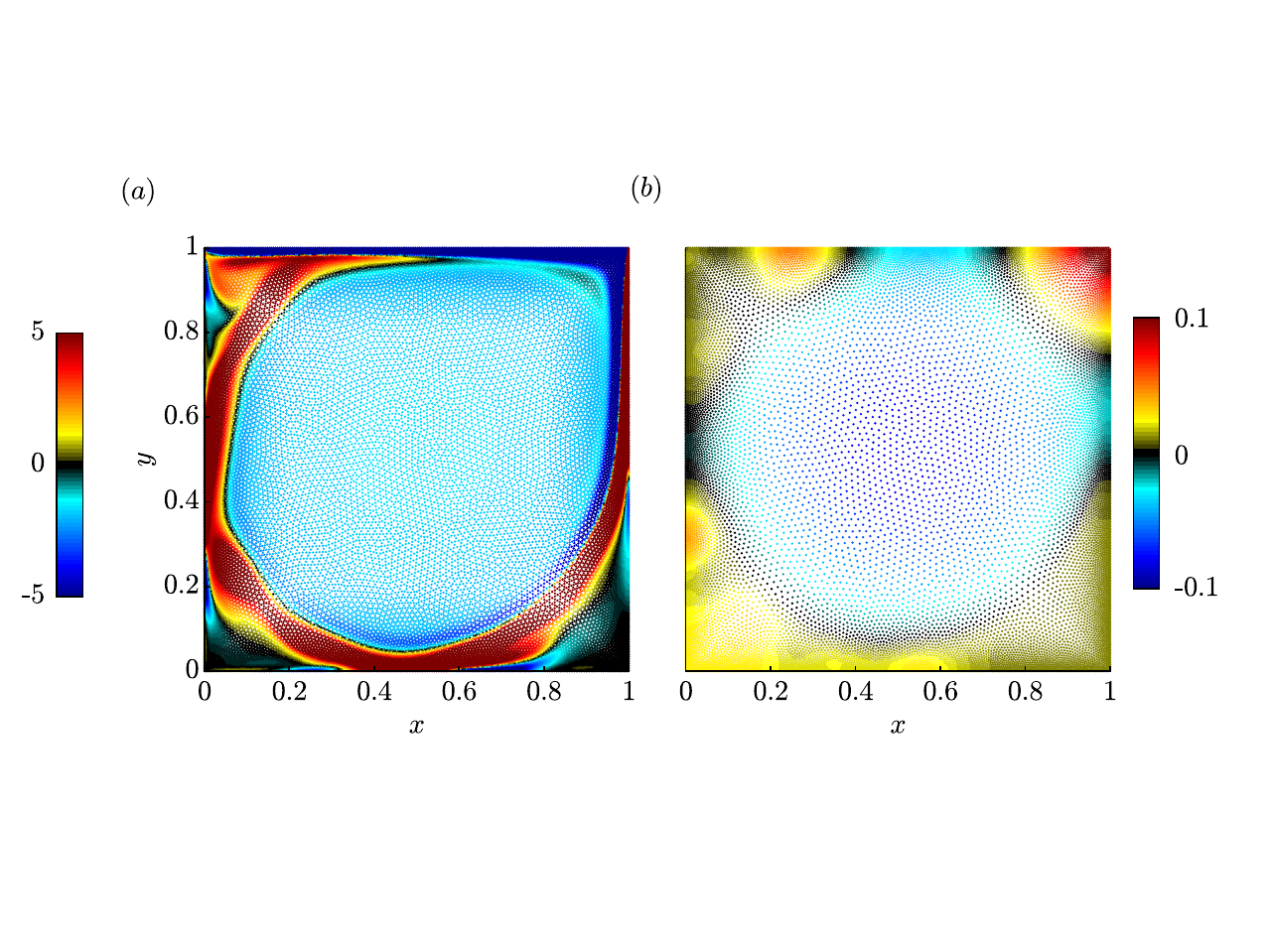}
\caption{Computational domain and solution for the lid-driven cavity at $\mathrm{Re}=10000$: (a) V-grid with $N= 42799\approx 207^2$ nodes, colored by vorticity; (b) P-grid with $M=14606\approx 121^2$ nodes, colored by pressure. 
}\label{mesh_cavity}
\end{figure}

 Figure \ref{mesh_cavity} shows the discretization of the unit square cavity domain, $\Omega=[0,1]\times[0,1]$, with $N=42799\approx 207^2$ and $M=14606\approx 121^2$. 
 The characteristic distance of the P-grid is 0.004 near the walls and averages at 0.008 for the whole domain. 
The flow inside the square cavity is driven by the motion of the top wall with unit velocity, $U_0=1$. No-slip boundary conditions are prescribed at all walls. 
The fluid is at rest at $t=0$.
 A time step $\Delta t = 0.00125$, corresponding to a CFL number around 0.7, is used in the computation. Results at Reynolds numbers, $\mathrm{Re}=\frac{U_0}{\nu}$, ranging from 100 to 10000 are investigated for comparison with the benchmark results by \citet{ghia1982high}.
    While other RBF-based codes often rely on hyperviscosity or other means of regularization, the present implementation runs stably for the entire range of Reynolds numbers.

\begin{figure}[hbt!]
\centering
\includegraphics[trim = 8mm 20mm 8mm 15mm, clip, width=0.8\textwidth]{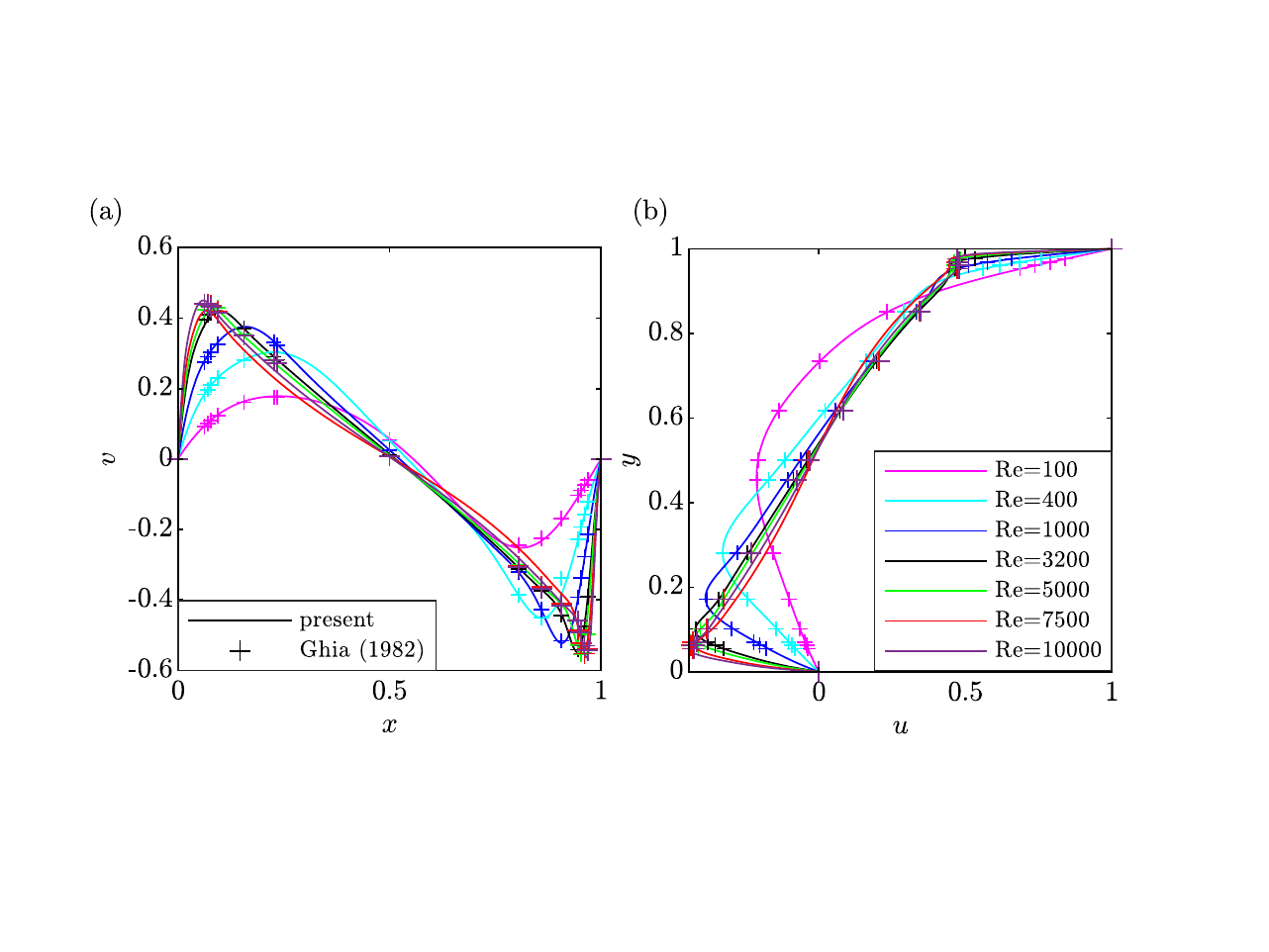}
\caption{Velocity profiles for the lid-driven cavity flow
at $\mathrm{Re}=100$ (magenta), $\mathrm{Re}=400$ (cyan), $\mathrm{Re}=1000$ (black), $\mathrm{Re}=3200$ (black), $\mathrm{Re}=5000$ (green) and $\mathrm{Re}=7500$ (red) through the: (a) horizontal centerline; (b) vertical centerline. Results at $\mathrm{Re}=10000$ (purple) are the time-averaged profiles.
The obtained results (solid) are compared to those of \citet{ghia1982high} (`+').}\label{velocity_centerline}
\end{figure}

  Figure \ref{velocity_centerline}(a) and (b) display the obtained velocity profiles through the horizontal and vertical centerlines of the cavity, respectively. The present results compare well with the benchmark data of \citet{ghia1982high}, who used $129^2$ 
    and $257^2$ nodes for Reynolds numbers smaller and larger or equal than 5000, respectively. After reaching steady-state, a periodic solution with frequency $f=0.63$ is found at $\mathrm{Re}=10000$. The corresponding curves in figure \ref{velocity_centerline} are therefore time-averaged velocity profiles.
The oscillation frequency varies in the literature, and the frequency obtained here falls well in the reported range \citep{auteri2002numerical,bruneau20062d,peng2003transition,tiesinga2002bifurcation}. After confirming the quantitative agreement of our results with the literature, we next examine the corresponding flow fields.

\begin{figure}[hbt!]
\centering
\includegraphics[trim = 10mm 20mm 5mm 20mm, clip, width=0.8\textwidth]{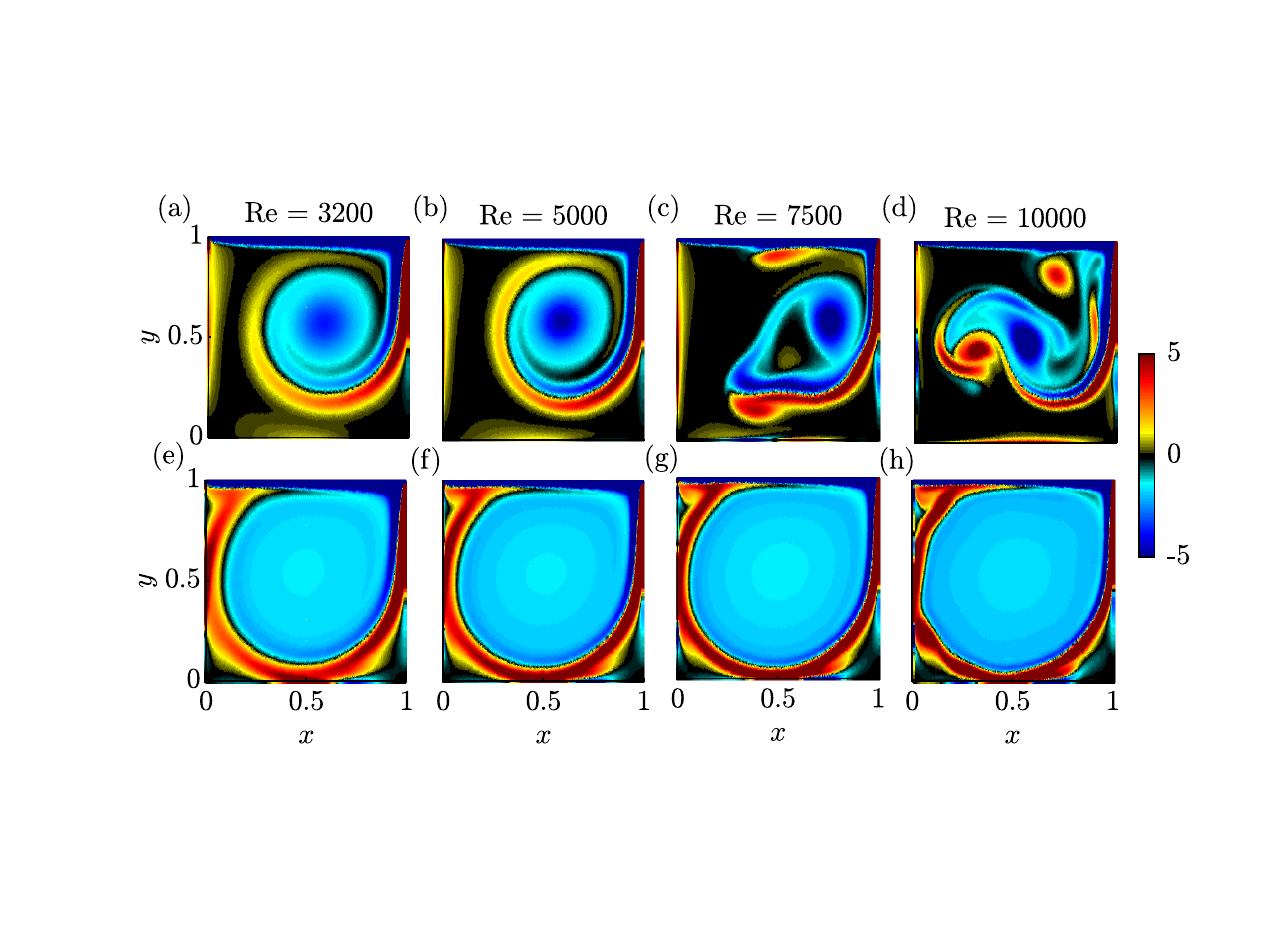}
\caption{Vorticity fields for $\mathrm{Re}=3200$ (a, e), $\mathrm{Re}=5000$ (b, f), $\mathrm{Re}=7500$ (c, g) and $\mathrm{Re}=10000$ (d, h) at: (a-d) $t=15$; (e-g) steady-state; (h) limit-cycle .
The vorticity vector is calculated as $\vb*{\omega}=\vb*{D}_x^{(V,V)}\vb*{v}-\vb*{D}_y^{(V,V)}\vb*{u}$. }\label{vorticity_evolution}
\end{figure}

   Figure \ref{vorticity_evolution} shows the vorticity fields for the four highest Reynolds numbers.
    For $\mathrm{Re}\leq 5000$, the primary vortices seen in figure  \ref{vorticity_evolution}(a,b) evolve into the steady-state solutions shown in \ref{vorticity_evolution}(e,f).
    Despite its more chaotic evolution, the flow at $\mathrm{Re}=7500$ also converges to a steady state.
It is generally observed that the higher the Reynolds numbers, the longer it takes to reach steady-state.
     For $\mathrm{Re}=10000$, the flow-field is doe not possess a steady-state solution, and we hence show an instantaneous state within the limit-cycle.
After confirming that the proposed scheme is well-suited for simulating this unsteady incompressible internal flow, 
 we next focus on the more challenging example of open flow over a cylinder.

\subsection{Cylinder flow} \label{cylinder flow}

 Wakes behind bluff bodies are an ubiquitous phenomenon in engineering, and the flow over a circular cylinder is often used as an unsteady benchmark problem for numerical methods.
RBF-based discretization facilitates easy local grid refinement near the solid boundaries and in the wake region.
    To the best knowledge of the authors, previous works on RBF for cylinder flows have exclusively used polar meshes to discretize the region around the cylinder \cite{ding2004simulation,javed2013hybrid,javed2014shape,shu2005computation,xie2021improved} (or have not reported the mesh topology).
 Here, we use scattered nodes, as described in \S \ref{Spatial Discretization} and previously shown in figure \ref{staggered}, throughout the entire domain.

\begin{figure}[hbt!]
\centering
\includegraphics[trim = 3mm 33mm 5mm 30mm, clip, width=0.85\textwidth]{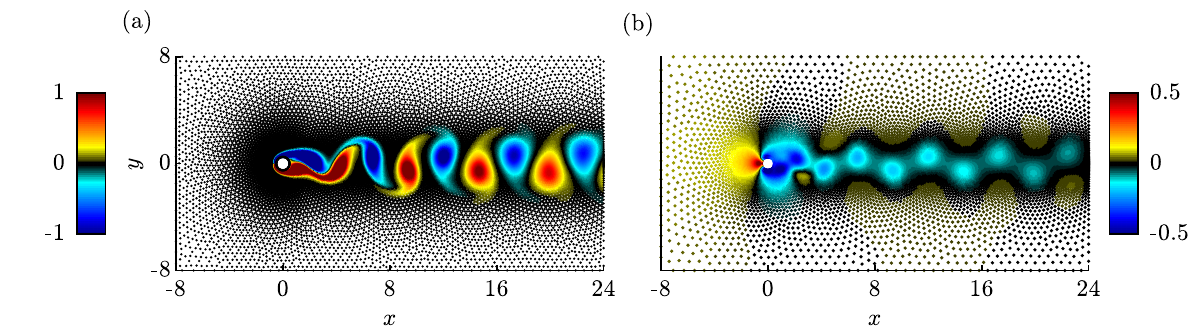}
\caption{Computational grids for cylinder flow: (a) V-grid with $N= 55671$ nodes showing $\omega$; (b) P-grid with $M=18647$ nodes showing $p$ at Re=100, respectively. }\label{mesh_cylinder}
\end{figure}

 Following \cite{calhoun2002cartesian,russell2003cartesian}, we take the computational domain $\Omega$ as the exterior of the cylinder $r\geq D/2 =0.5$ within the rectangle $-8\leq x\leq 24,\, -8\leq y\leq 8$. Local grid refinement is used to resolve the regions near the cylinder and the wake. 
  Figure \ref{mesh_cylinder}(a,b) show the V-grid with $N= 55671$ and P-grid with $M = 18647$, respectively. The characteristic distances of the P-grid are $\Delta r^{(p)}=0.03$ near the cylinder, 0.04 on the wake centerline, $x > 0.5, y = 0$, and averaged at 0.123 in the whole domain. 
 The inflow is uniform with  $U_{\infty}=1,\, V_{\infty}=0$, symmetric boundary conditions with $v=\partial u/\partial y =0$ are applied at the transverse boundaries, a no-slip condition is prescribed on the cylinder, and a stress-free outflow condition, $-p\vb*{n}+\frac{1}{\mathrm{Re}} \nabla \vb*{U}\cdot\vb*{n}=0$, where $\vb*{n}=[1,0]^T$, is prescribed at the outlet.
 The flow field is initialized with $\vb*{U}_0=[\vb*{1},\vb*{0}]^T$. 
    The time step is $\Delta t=0.005$, corresponding to a CFL number of 0.67. Two Reynolds numbers, $\mathrm{Re}= \frac{U_{\infty} D}{\nu}=$100 and 200, are considered.


\begin{figure}[hbt!]
\centering
\includegraphics[trim = 15mm 20mm 5mm 16mm, clip, width=0.8\textwidth]{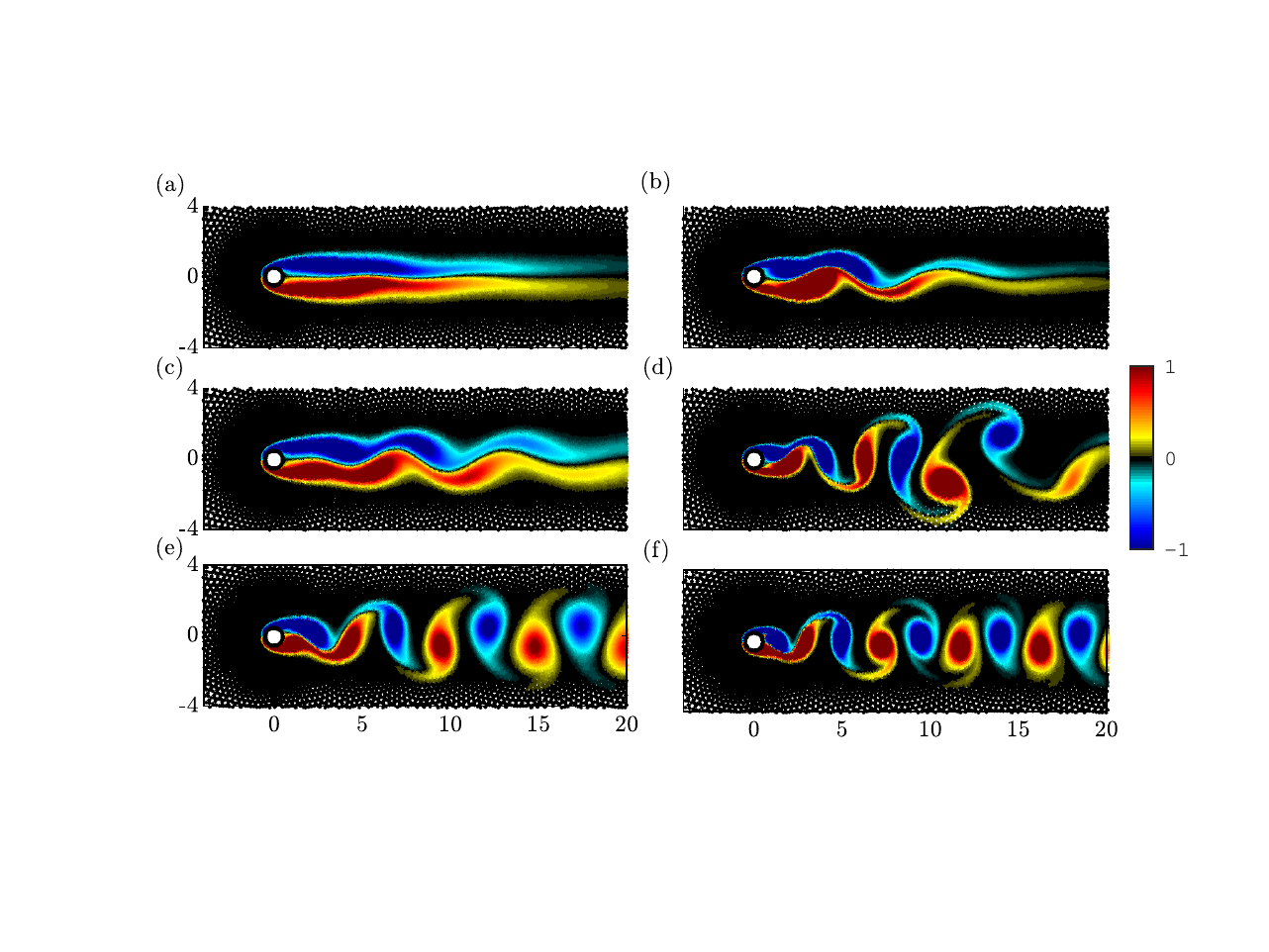}
\caption{
Instantaneous vorticity fields for $\mathrm{Re}=100$ (a, c, e), and $\mathrm{Re}=200$ (b, d, h) at: (a, b) $t=25$; (c, d) $t=40$; (e, f) limit-cycle .
} \label{vortex_shedding}
\end{figure}

    Figure \ref{vortex_shedding} shows the vorticity field at three different time instances for both Reynolds numbers.
    For both cases, these three time instances capture the gradual evolution of the unstable wake flow through a transient that finally results in the limit-cycle oscillation. This limit-cycle oscillation is the well-known periodic von K{\'a}rm{\'a}n vortex street seen in figures \ref{vortex_shedding}(e) and \ref{vortex_shedding}(f).

\begin{figure}[hbt!]
\centering
\includegraphics[trim = 0mm 2mm 0mm 42mm, clip, width=0.7\textwidth]{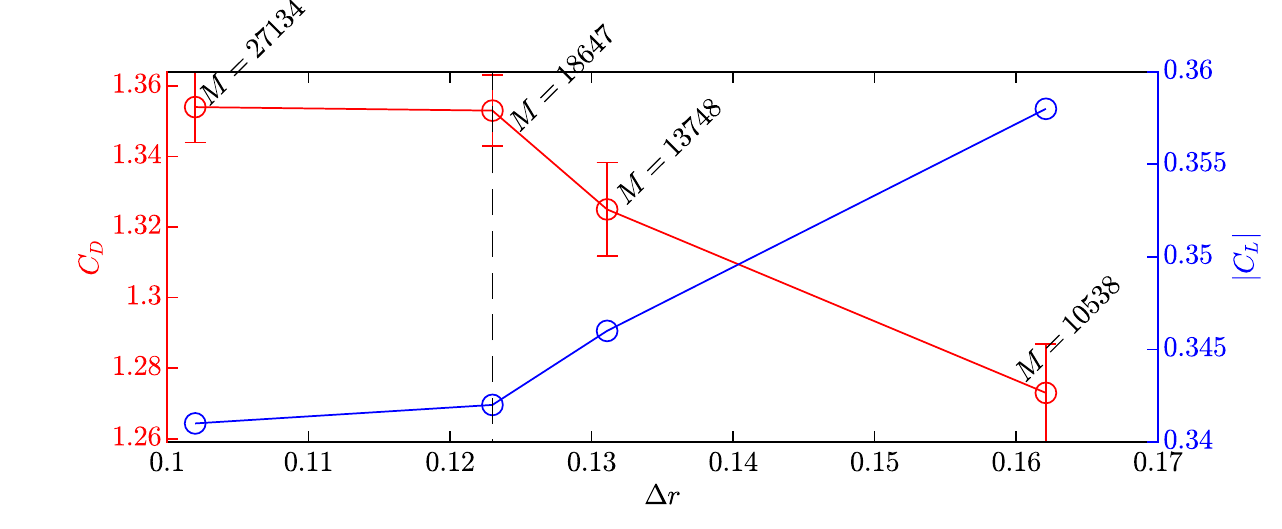}
\caption{
Drag (red) and lift (blue) coefficients for cylinder flow at $\mathrm{Re}=100$ for different grid resolutions. {The vertical dashed line indicates the reference P-grid resolution $\Delta r=0.123$ used in \S \ref{cylinder flow}.}}\label{coeff_Re_100}
\end{figure}
We next quantify these results by examining the drag and lift coefficients on the limit cycle, defined as
\begin{align}
    C_D = \frac{F_D}{\frac{1}{2}\rho U_{\infty}^2 D},\qquad C_L = \frac{F_L}{\frac{1}{2}\rho U_{\infty}^2 D},
\end{align}
respectively. The drag, $F_D$, and lift, $F_L$, forces are computed by integrating the pressure and wall-shear over the cylindrical surface using the Simpson’s rule. The grid dependence of the lift and drag coefficients is investigated in figure \ref{coeff_Re_100} for $\mathrm{Re}=100$. 
The four grids are obtained by varying the initial edge length for {\tt DistMesh} from $0.025$ to $0.04$.
The differences in the drag and lift coefficients between the current (vertical dashed line) and the densest grid are $0.07\%$ and $0.29\%$, respectively. This indicates that the present grid resolution is sufficient.

\begin{table}[!htb]
      \centering
        \begin{tabular}{|c|c|c|c||c|c|c| }
      \cline{2-7}
  \multicolumn{1}{c|}{}   &   \multicolumn{3}{ c|| }{$\mathrm{Re}=100$} &  \multicolumn{3}{ c| }{$\mathrm{Re}=200$}\\ \cline{2-7}
    \multicolumn{1}{c|}{}    & $C_D$ & $C_L$ & $\mathrm{St}$& $C_D$ &  $C_L$ & $\mathrm{St}$\\
 \hline \hline
 \citet{braza1986numerical} & $1.364\pm 0.015$ & $\pm 0.25$ & - & $1.40\pm 0.05$ & $\pm 0.75$ & -\\ 
 \hline
 \citet{liu1998preconditioned} & $1.350\pm 0.012$ & $\pm 0.339$ & 0.164 & $1.31\pm 0.049$ & $\pm 0.69$ & 0.192\\
 \hline
 \citet{calhoun2002cartesian} & $1.330\pm 0.014$ & $\pm 0.298$ & 0.175  & $1.17\pm 0.058$ & $\pm 0.67$ & 0.202 \\
 \hline
 \citet{russell2003cartesian} & $1.38\pm 0.007$ & $\pm 0.300$ & 0.169 & $1.29\pm 0.022$ & $\pm 0.50$ & 0.195 \\
 \hline
    \citet{ding2004simulation} & $1.325\pm 0.008$ & $\pm 0.28$ & 0.164 & $1.327\pm 0.045$ & $\pm 0.60$ & 0.196 \\
 \hline
   \citet{shu2005computation} & $1.362\pm 0.010$ & $\pm 0.32$ & 0.166 & $1.352\pm 0.049$ & $\pm 0.62$ & 0.192 \\
 \hline
    \citet{shahane2021high} & $1.354\pm 0.009$ & $\pm 0.333$ & 0.166 & $1.364\pm 0.045$ & $\pm 0.690$ & 0.197 \\
 \hline
 Present & 1.353 $\pm$ 0.010 & $\pm$ 0.342 &  0.171& 1.382 $\pm$ 0.046  & $\pm$ 0.683 &  0.201\\ 
 \hline
        \end{tabular}
        \caption{Summary of published results for drag, lift and the fundamental vortex shedding frequency the flow behind a cylinder at $\mathrm{Re}=100$ and $\mathrm{Re}=200$, and comparison to current results.}
    \label{results_cylinder}
\end{table}

 Table \ref{results_cylinder} compares previous results for both Reynolds numbers to the values obtained here. The vortex shedding frequency is reported in terms of the Strouhal number, $\mathrm{St}=f D/U_{\infty}$.
It can be seen that our results fall well within the range of the previous results, which vary by as much as $16\%$ at $\mathrm{Re}=200$.
 Out of the reported literature, the work of  \citet{shahane2021high} is most closely related to the present study. \citet{shahane2021high} report results for their collocated PHS-RBF method for two different grid resolutions. 
Their results on the finer grid are reported in table \ref{results_cylinder}.
The staggered-grid arrangement used here utilizes around $65 \%$ fewer nodes on the $V-$grid, or $85\%$ on the $P-$grid, and an about one-third smaller RBF stencil.

\section{Summary and discussion} \label{conclusion}

The main objective of this work was to advance the state-of-the-art in RBF-based discretizations to enable future high-fidelity simulations of engineering or natural flows in complex domains. To this end, a semi-implicit fractional-step algorithm for the time-dependent Navier-Stokes equations that uses PHS+poly RBF-FDs to approximate spatial derivatives on scattered nodes was presented. The algorithm achieves high accuracy and stability by staggering the pressure at the vertices and the velocities at the face centers of unstructured triangular meshes. A systematic study has shown that the parameter combination $(n,m,q)=(28,7,3)$ yields a good balance between accuracy and stencil size. In particular, this parameter triplet ensures high accuracy up to a proposed maximum modified wavenumber of $\tilde{k} \Delta r =\sqrt{2}$, that is, the same `spectral-like resolution' of higher-order Pad\'e-type compact finite differences \citep{lele1992compact}. For a stencil size $n=28$, it was found that increasing the polynomial degree beyond $q=3$ does not improve accuracy and that a PHS exponent of $m=7$ yields the smallest error within the considered parameter space. A two-dimensional modified wavenumber analysis has further shown that the error between the computed derivatives and the spectral limit does not exceed $1.5\%$ at $\tilde{k} \Delta r =\sqrt{2}$ for varying wave angles, hence demonstrating high-order accuracy in the presence of grid non-homogeneity. The incompressible Navier-Stokes solver is demonstrated on two benchmark problems, lid-driven cavity, and cylinder flow. Over a range of Reynolds numbers, the results compare well with data reported in previous works at a comparably low computational cost.

  

Since the pioneering work by \citet{hardy1971multiquadric}, RBF discretizations have been applied to a large variety of mostly canonical problems and have been continuously improved and specialized. Take as examples most relevant to this contribution the introduction of RBF-FDs \citep{tolstykh2000using}, PHS+poly RBFs \citep{flyer2016enhancing}, and, most recently, the use of PHD+poly RBFs to solve the incompressible Navier-Stokes equations by \citet{shahane2021high}. Despite the impressive conceptual and algorithmic progress, the technology transition to a general-purpose CFD tool for physical exploration or engineering applications has, arguably, not been yet accomplished in the same way as for FD, FV, finite element, and DG methods. The present contribution should be seen as a continuation of the works referenced above with this goal. The natural next steps along these lines are hence the extension to 3-D and demonstration on non-canonical geophysical or engineering flows. A conceptual advantage of RBF-FDs, which it shares with classical FDs, is that they can directly be used in the strong formulation. This makes the construction of sparse differentiation operators, and therefore global Jacobians for hydrodynamic stability analysis, particularly easy. The latter application is another direction of our current research.
  




  


\section*{Acknowledgments}

 The authors would like to thank Bengt Fornberg for sharing his insights. We gratefully acknowledge support by the National Science Foundation under Grant No. CBET-1953999 (PM Ron Joslin).



\setcounter{equation}{0}
\renewcommand{\theequation}{{\rm A}.\arabic{equation}}

\appendix

\section{Modified wavenumber diagrams for $q=4$}

\begin{figure}[hbt!]
\centering
\includegraphics[trim = 0mm 0mm 0mm 0mm, clip, width=0.8\textwidth]{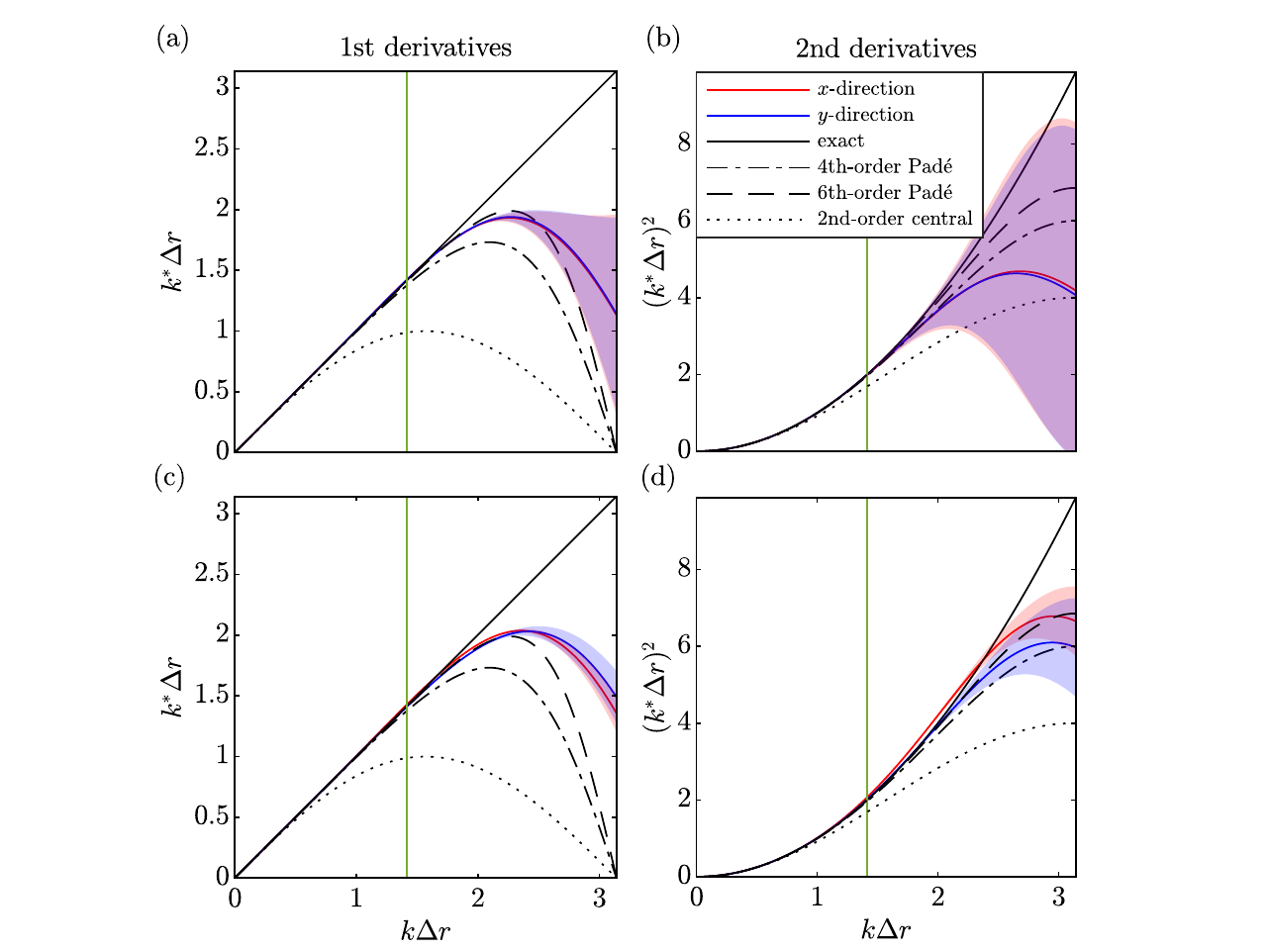}
\caption{Modified wavenumber diagrams for $(n,m,q)=(28,7,4)$: (a,b) same as figure \ref{mw_V2V} for $\vb*{D}^{(V,V)}$; (c,d): same as figure \ref{mw_P2P} for $\vb*{D}^{(P,P)}$.}\label{mw_q4}
\end{figure}

Panels \ref{mw_q4}(a,b) show the modified wavenumber diagrams for the differentiation matrices $\vb*{D}^{(V,V)}$ obtained using the combination $(n,m,q)=(28,7,4)$. The results are visually indistinguishable from those in figure \ref{mw_V2V}, indicating that increasing the polynomial degree, $q$, beyond 3 does not improve the accuracy for a fixed RBF stencil size, here $n=28$.
The corresponding diagram for the $(P,P)-$grid is shown in panels \ref{mw_q4}(c,d). In the same way as for the $(V,V)-$grid, the modified wavenumber curves in figures \ref{mw_P2P} and \ref{mw_q4}(c,d) are nearly identical.

\section{Relative errors for V-to-P and P-to-V grid derivatives} \label{V2P_P2V}

\begin{figure}[hbt!]
\centering
\includegraphics[trim = 0mm 0mm 0mm 0mm, clip, width=1\textwidth]{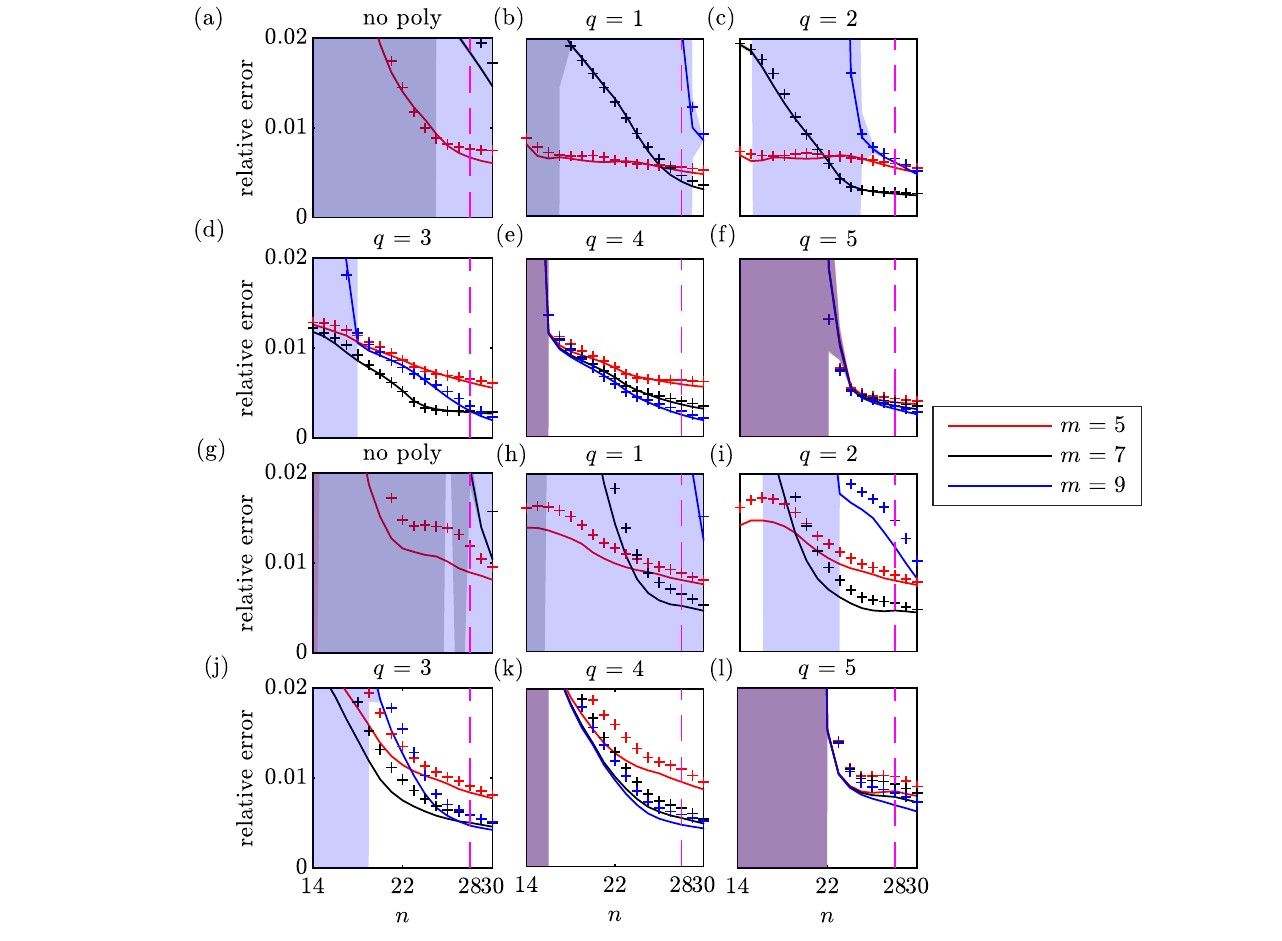}
\caption{
Relative errors for staggered nodes: (a-f) same as figure \ref{error_dx_V2V} for $\vb*{D}^{(V,P)}_{x}$; (g-l) same as figure \ref{error_dx_V2V} for $\vb*{D}^{(P,V)}_{x}$. The results for the $y$-derivatives are shown as `+'.}\label{error_dx}
\end{figure}

The staggered node layout introduced in \S \ref{Spatial Discretization} requires differentiation matrices that operate between the $V$- and $P$-grids (see equations (\ref{divergence}) and (\ref{velocity_next_timestep}) of the fractional step algorithm). Panels \ref{error_dx}(a-f) show the relative errors for the differentiation matrix $\vb*{D}^{(V,P)}$, which operates on the $(V,P)$-grid shown in figure \ref{mesh_neighbor}(c) for varying $n$, $m$ and $q$. Similar trends as for the collocated $(V,V)$- and $(P,P)$-grids are found. For the same reasons discussed in \S \ref{error_analysis}, and for consistency with the differentiation operators on the collocated grids, we proceed with the same parameter combination of  $(n,m,q)=(28,7,3)$. The same parameter study is repeated for the differentiation operator $\vb*{D}^{(P,V)}$ in panels \ref{error_dx}(g-l) with a comparable outcome and the same conclusion.

\section{Comparison with RBF-QR and RBF-GA}\label{RBF_comparison}

\begin{figure}[hbt!]
\centering
\includegraphics[trim = 10mm 20mm 10mm 15mm, clip, width=0.7\textwidth]{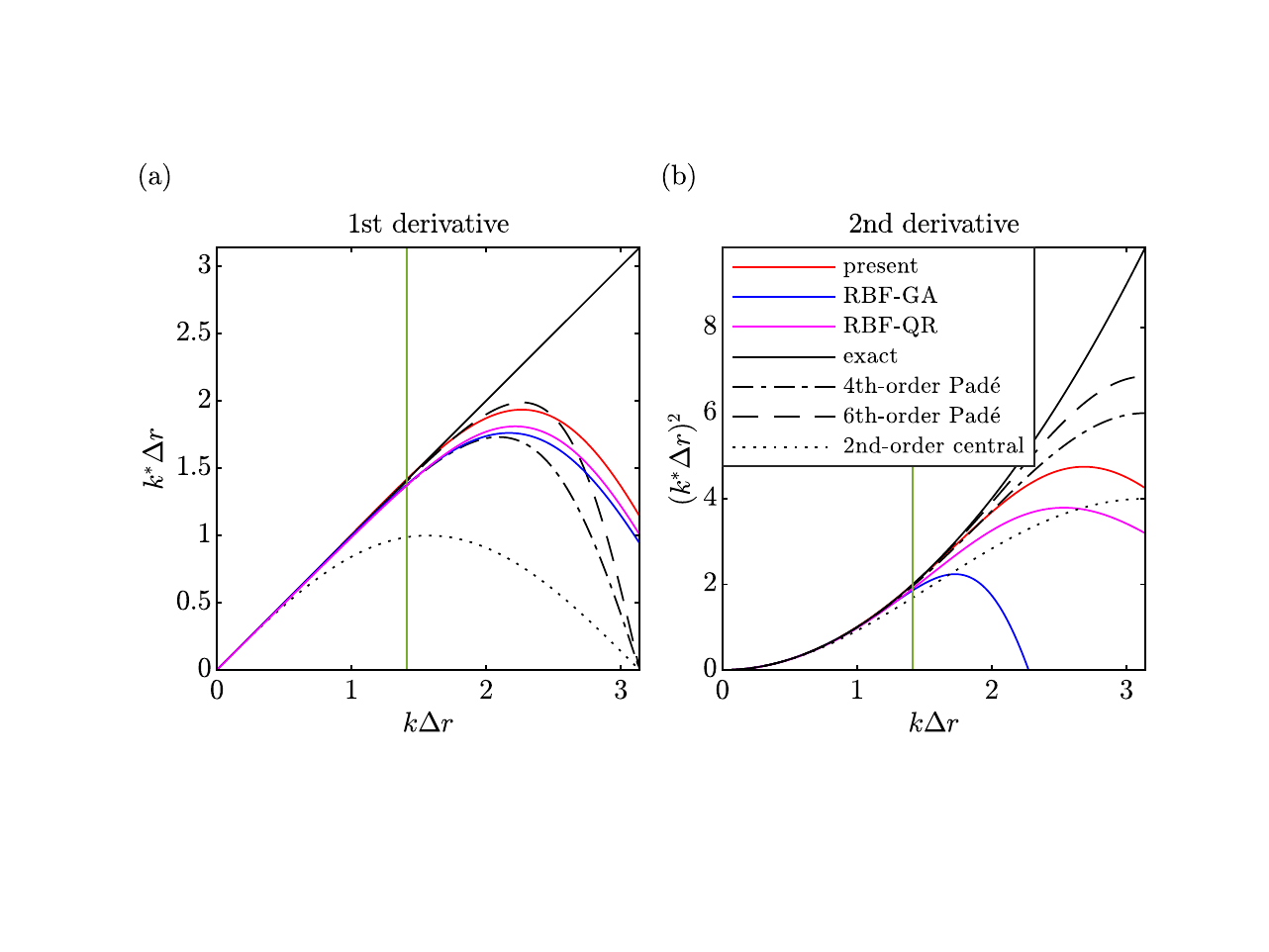}
\caption{Comparison with RBF-QR and RBF-GA on the (V,V)-grid with $n=28$: modified wavenumber diagrams for the (a)  first derivative; (b) second derivative. The recommended maximum modified wavenumber of $k\Delta r=\sqrt{2}$ is shown in green. }\label{mw_V2V_comparison}
\end{figure}

Figure \ref{mw_V2V_comparison} compares the accuracy of different RBF-FD methods for a fixed stencil size of $n=28$.
The comparison is conducted on the $(V,V)$-grid, on which both the first and second derivatives are required.
The accuracy and numerical stability of the RBF-QR and RBF-GA methods depend on their shape parameters.
The corresponding values of $\epsilon_{\text{QR}}=2.5$ and $\epsilon_{\text{GA}}=...$ are taken from the standard references by \citet{fornberg2011stabilization} and \citet{bollig2012solution}, respectively. 
It can be seen that all three RBF-FD variants perform very well up to our recommended maximum modified wavenumber of $k\Delta r=\sqrt{2}$ for both the first and second derivatives.
For fewer points per wavelength, that is $k\Delta r\gtrsim \sqrt{2}$, the proposed PHS+poly variant with $(n,m,q)=(28,7,3)$ stays closer to the spectral limit, in particular for the second derivative.
With the caveat that the shape parameters of the RBF-QR and RBF-GA variants are not optimized for this specific node distribution, the results show that the proposed PHS+poly discretization is highly competitive. 
An advantage of PHS+poly, arguably, is that the optimization of its integer parameters, $m$ and $q$ as demonstrated in \S \ref{RBF}, for a fixed stencil size is more straight forward than finding the optimal value of a continuous shape parameter.

\section{Time stepping accuracy} \label{timestep}

\begin{figure}[hbt!]
\centering
\includegraphics[trim = 0mm 2mm 0mm 0mm, clip, width=0.6\textwidth]{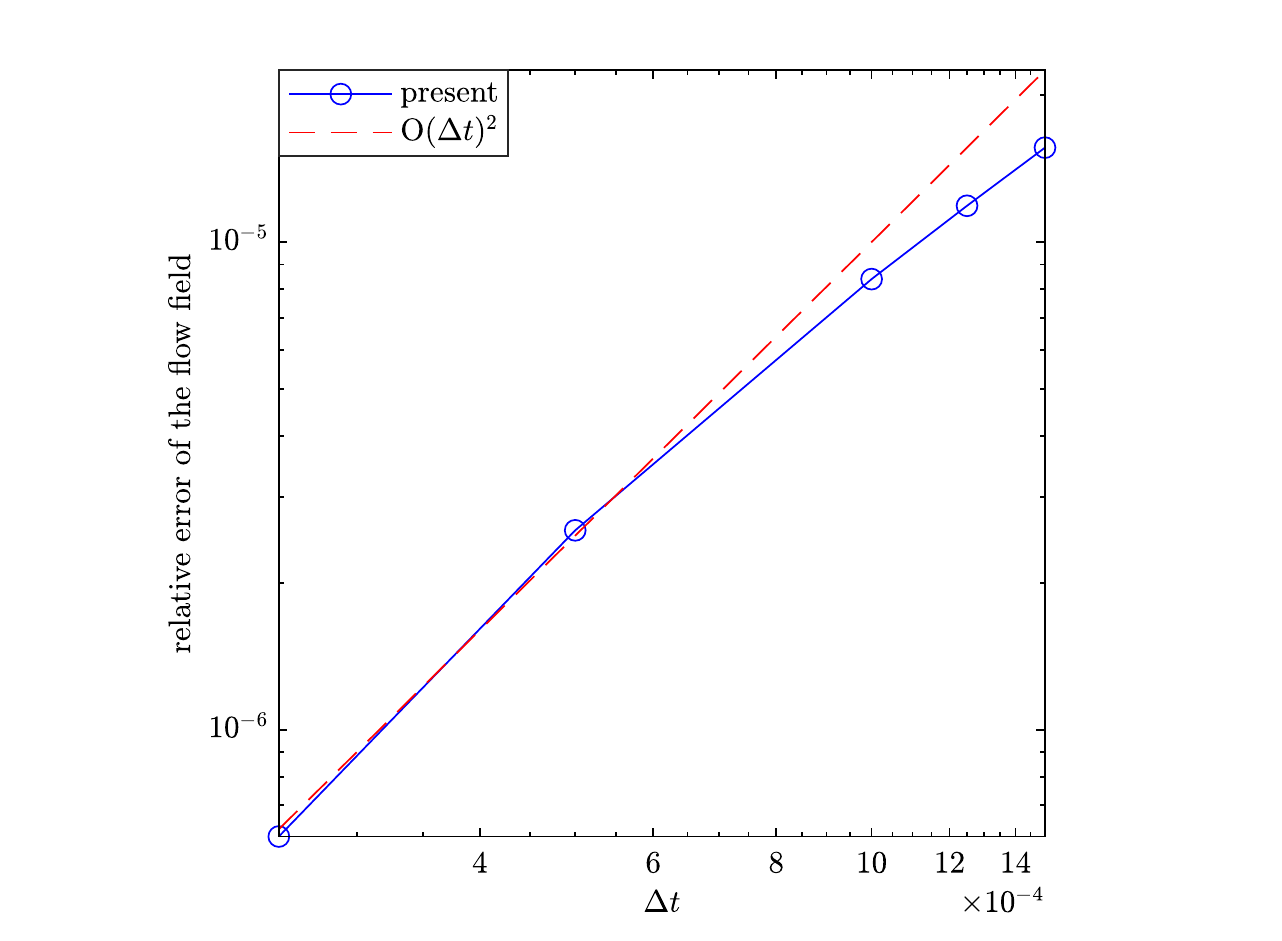}
\caption{
Relative error of the flow field at different time steps.}\label{timestep_accuracy}
\end{figure}

Figure \ref{timestep_accuracy} shows the relative error of the flow field simulated using different time steps. As a test case, we use the lid-driven cavity flow at $\mathrm{Re}=1000$ and compute the error for the steady-state solution, see \S \ref{Lid-Driven cavity flows}. The close match between the theoretical curve and the simulation error for small $\Delta t$ confirms that the algorithm possesses an order of accuracy of $O(\Delta t)^2$. This is consistent with the numerical scheme described in \S \ref{Spatial Discretization}.


\newpage

 \begin{thebibliography}{00}




\bibitem[Auteri \emph{et al.}(2021)]{auteri2002numerical}
F. Auteri, N. Parolini, L. Quartapelle,\, {Numerical investigation on the stability of singular driven cavity flow},\,
J. Comput. Phys. 183 (1)(2002) 1--25.



\bibitem[Bartwal \emph{et al.}(2021)]{bartwal2021application}
N. Bartwal, S. Shahane, S. Roy, S. P. Vanka,\, {Application of a High Order Accurate Meshless Method to Solution of Heat Conduction in Complex Geometries},\,
arXiv preprint arXiv:2106.08535. (2021).



\bibitem[Bayona(2019a)]{bayona2019insight}
V. Bayona,\, {An insight into RBF-FD approximations augmented with polynomials},\,
Comput. Math. Appl. 77 (9)(2019) 2337--2353.


\bibitem[Bayona(2019b)]{bayona2019comparison}
V. Bayona,\, {Comparison of moving least squares and RBF+ poly for interpolation and derivative approximation},\,
J. Sci. Comput. 81 (1)(2019) 486--512.


\bibitem[Bayona \emph{et al.}(2017)]{bayona2017role}
V. Bayona, N. Flyer, B. Fornberg, G.A. Barnett,\, {On the role of polynomials in RBF-FD approximations: II. Numerical solution of elliptic PDEs},\,
J. Comput. Phys. 332 (2017) 257--273.


\bibitem[Bayona \emph{et al.}(2019)]{bayona2019role}
V. Bayona, N. Flyer, B. Fornberg,\, {On the role of polynomials in RBF-FD approximations: III. Behavior near domain boundaries},\,
J. Comput. Phys. 380 (2019) 378--399.




\bibitem[Bollig \emph{et al.}(2012)]{bollig2012solution}
E. F. Bollig, N. Flyer, G. Erlebacher,\, {Solution to PDEs using radial basis function finite-differences (RBF-FD) on multiple GPUs},\,
J. Comput. Phys. 231 (21)(2012) 7133--7151.



\bibitem[Braza \emph{et al.}(1986)]{braza1986numerical}
M. Braza, P. Chassaing,  H. Ha Minh,\, {Numerical study and physical analysis of the pressure and velocity fields in the near wake of a circular cylinder},\,
J.~Fluid Mech. 165 (1986) 79--130.




\bibitem[Bruneau and Saad(1986)]{bruneau20062d}
C. H. Bruneau, M. Saad,\, {The 2D lid-driven cavity problem revisited},\,
Comput. Fluids. 35 (3)(2006) 326--348.



\bibitem[Calhoun(2002)]{calhoun2002cartesian}
D. Calhoun,\, {A Cartesian grid method for solving the two-dimensional streamfunction-vorticity equations in irregular regions},\,
J. Comput. Phys. 176 (2)(2002) 231--275.

\bibitem[Carlson and Foley(1991)]{carlson1991parameter}
R.E. Carlson, T.A. Foley,\, {The parameter R2 in multiquadric interpolation},\,
Comput. Math. Appl. 21 (9)(1991) 29--42.


\bibitem[Chinchapatnam \emph{et al.}(2016)]{chinchapatnam2009compact}
P.P. Chinchapatnam, K. Djidjeli, P.B. Nair, M. Tan,\, {A compact RBF-FD based meshless method for the incompressible Navier—Stokes equations},\,
Proc. IME. M-J Eng. Marit. Environ. 316 (3)(2016) 275--290.






\bibitem[Colonius \emph{et al.}(1995)]{colonius1997sound}
T. Colonius, S.K. Lele, P. Moin,\, {Sound generation in a mixing layer},\,
J. Fluid Mech. 330 (1997) 375--409.





\bibitem[Ding \emph{et al.}(2004)]{ding2004simulation}
H. Ding, C. shu, K.S. Yeo, D. Xu,\, {Simulation of incompressible viscous flows past a circular cylinder by hybrid FD scheme and meshless least square-based finite difference method},\,
Comput. Methods Appl. Mech. Engrg. 193 (9--11)(2004) 727--744.



\bibitem[Ding \emph{et al.}(2006)]{ding2006numerical}
H. Ding, C. shu, K.S. Yeo, D. Xu,\, {Numerical computation of three-dimensional incompressible viscous flows in the primitive variable form by local multiquadric differential quadrature method},\,
Comput. Methods Appl. Mech. Engrg. 195 (7--8)(2006) 516--533.


\bibitem[Fasshauer and Zhang(2007)]{fasshauer2007choosing}
G.E. Fasshauer, J.G. Zhang,\, {On choosing “optimal” shape parameters for RBF approximation},\,
Numer. Algorithms. 45 (1--4)(2007) 345--368.


\bibitem[Flyer \emph{et al.}(2016)]{flyer2016enhancing}
N. Flyer, G.A. Barnett, L.J. Wicker,\, {Enhancing finite differences with radial basis functions: experiments on the Navier--Stokes equations},\,
J. Comput. Phys. 316 (2016) 39--62.


\bibitem[Flyer and Fornberg (2011)]{flyer2011radial}
N. Flyer, B. Fornberg,\, {Radial basis functions: Developments and applications to planetary scale flows},\,
Comput. \& Fluids  46 (1)(2011) 23--32.


\bibitem[Flyer \emph{et al.}(2016)]{flyer2016role}
N. Flyer, B. Fornberg, V. Bayona, G.A. Barnett,\, {On the role of polynomials in RBF-FD approximations: I. Interpolation and accuracy},\,
J. Comput. Phys. 321 (2016) 21--38.



\bibitem[Flyer and Lehto(2010)]{flyer2010rotational}
N. Flyer, E. Lehto,\, {Rotational transport on a sphere: Local node refinement with radial basis functions},\,
J. Comput. Phys. 229 (6)(2010) 1954--1969.



\bibitem[Flyer \emph{et al.}(2012)]{flyer2012guide}
N. Flyer, E. Lehto, S. Blaise, G. Wright,\, {A guide to RBF-generated finite differences for nonlinear transport: Shallow water simulations on a sphere},\,
J. Comput. Phys. 231 (11)(2012) 4078--4095.



\bibitem[Flyer and Wright(2007)]{flyer2007transport}
N. Flyer, G. B. Wright,\, {Transport schemes on a sphere using radial basis functions},\,
J. Comput. Phys. 226 (1)(2007) 1059--1084.

\bibitem[Flyer and Wright(2009)]{flyer2009radial}
N. Flyer, G. B. Wright,\, {A radial basis function method for the shallow water equations on a sphere},\,
Proc. Math. Phys. Eng. Sci. 465 (2106)(2009) 1949--1976.


\bibitem[Fornberg \emph{et al.}(2002)]{fornberg2002observations}
B. Fornberg, T. A. Driscoll, G. Wright, R. Charles,\, {Observations on the behavior of radial basis function approximations near boundaries},\,
Comput. Math. Appl. 43 (3-5)(2002) 473--490.


\bibitem[Fornberg and Flyer(2015)]{fornberg2015solving}
B. Fornberg, N. Flyer,\, {Solving PDEs with radial basis functions},\,
Acta Numer. 24 (2015) 215--258.

\bibitem[Fornberg and Flyer(2011)]{fornberg2011stable}
B. Fornberg, E. Larsson, N. Flyer,\, {Stable computations with Gaussian radial basis functions},\,
SIAM J. Sci. Comput. 33 (2)(2011) 869--892.

\bibitem[Fornberg and Letho(2011)]{fornberg2011stabilization}
B. Fornberg, E. Lehto,\, {Stabilization of RBF-generated finite difference methods for convective PDEs},\,
J. Comput. Phys. 230 (6)(2011) 2270--2285.


\bibitem[Fornberg \emph{et al.}(2013)]{fornberg2013stable}
B. Fornberg, E. Lehto, C. Powell,\, {Stable calculation of Gaussian-based RBF-FD stencils},\,
Comput. Math. Appl. 65 (4)(2013) 627--637.


\bibitem[Fornberg and Piret(2008)]{fornberg2008stable}
B. Fornberg, C. Piret,\, {A stable algorithm for flat radial basis functions on a sphere},\,
SIAM J. Sci. Comput. 30 (1)(2008) 60--80.

\bibitem[Fornberg and Wright(2004)]{fornberg2004stable}
B. Fornberg, G. Wright,\, {Stable computation of multiquadric interpolants for all values of the shape parameter},\,
Comput. Math. Appl. 48 (5--6)(2004) 853--867.








\bibitem[Franke (1982)]{franke1982scattered}
R. Franke,\, {Scattered data interpolation: tests of some methods},\,
Math. Comp. 38 (1982) 181--200.

\bibitem[Ghia \emph{et al.}(1982)]{ghia1982high}
U. Ghia, K.N. Ghia, C.T. Shin,\, {High-Re solutions for incompressible flow using the Navier-Stokes equations and a multigrid method},\,
J. Comput. Phys. 48 (3)(1982) 387--411.


\bibitem[Gunderman \emph{et al.}(2020)]{gunderman2020transport}
D. Gunderman, N. Flyer, B. Fornberg,\, {Transport schemes in spherical geometries using spline-based RBF-FD with polynomials},\,
J. Comput. Phys. 408 (2020) 109256.



\bibitem[Harlow and Welch(1965)]{harlow1965numerical}
F. H. Harlow, J. E. Welch,\, {Numerical calculation of time-dependent viscous incompressible flow of fluid with free surface},\,
Phys. Fluids. 8 (12)(1965) 2182--2189.


\bibitem[Hardy(1971)]{hardy1971multiquadric}
R. L. Hardy,\, {Multiquadric equations of topography and other irregular surfaces},\,
J. Geophys. Res. 76 (8)(1971) 1905--1915.





\bibitem[Iske (2003a)]{iske2003approximation}
A. Iske,\, {On the approximation order and numerical stability of local Lagrange interpolation by polyharmonic splines, in: W. Haussmann, K. Jetter, M. Reimer, J. St{\"o}ckler (Eds.)},\,
Modern developments in multivariate approximation, in: International Series of Numerical Mathematics, Birkh{\''a}user, Basel. (2003) 153--165.


\bibitem[Iske (2003b)]{iske2003radial}
A. Iske,\, {Radial basis functions: basics, advanced topics and meshfree methods for transport problems},\,
Rend. Sem. Mat. Univ. Pol. (Torino) 61 (3)(2003) 247--285.



\bibitem[Javed \emph{et al.}(2013)]{javed2013hybrid}
A. Javed, K. Djidjeli, J.T. Xing, S.J. Cox,\, {A hybrid mesh free local RBF-Cartesian FD scheme for incompressible flow around solid bodies},\,
Int. J. Math. Comput. Phys. Elect. Comput. Eng. 7 (2013) 957--966.


\bibitem[Javed \emph{et al.}(2014)]{javed2014shape}
A. Javed, K. Djidjeli, J.T. Xing,\, {Shape adaptive RBF-FD implicit scheme for incompressible viscous Navier--Strokes equations},\,
Comput. \& Fluids.  89 (2014) 38--52.


\bibitem[Joslin \emph{et al.}(1990a)]{joslin1993spatial}
R. D. Joslin, C.L. Streett, C.-L. Chang,\, {Spatial direct numerical simulation of boundary-layer transition mechanisms: Validation of PSE theory},\,
Theor. Comput. Fluid Dyn. 4 (6)(1993) 271--288.

\bibitem[Kansa(1990a)]{kansa1990multiquadrics1}
E. J. Kansa,\, {Multiquadrics—A scattered data approximation scheme with applications to computational fluid-dynamics—I surface approximations and partial derivative estimates},\,
Comput. Math. Appl.  19 (8-9)(1990) 127--145.



\bibitem[Kansa(1990b)]{kansa1990multiquadrics2}
E. J. Kansa,\, {Multiquadrics—A scattered data approximation scheme with applications to computational fluid-dynamics—II solutions to parabolic, hyperbolic and elliptic partial differential equations},\,
Comput. Math. Appl.  19 (8-9)(1990) 147--161.


\bibitem[Kansa and Hon(2000)]{kansa2000circumventing}
E. J. Kansa, Y. C. Hon\, {Circumventing the ill-conditioning problem with multiquadric radial basis functions: applications to elliptic partial differential equations},\,
Comput. Math. Appl.  39 (7-8)(2000) 123--137.

\bibitem[Kim and Moin(1985)]{kim1985application}
J. Kim, P. Moin,\, {Application of a fractional-step method to incompressible Navier-Stokes equations},\,
J. Comput. Phys. 59 (2)(1985) 308--323.



\bibitem[Kloker(1997)]{kloker1997robust}
M. J. Kloker,\, {A robust high-resolution split-type compact FD scheme for spatial direct numerical simulation of boundary-layer transition},\,
Appl. Sci. Res. 59 (4)(1997) 353--377.







\bibitem[Larsson \emph{et al.}(2013)]{larsson2013stable}
E. Larsson, E. Lehto, A. Heryudono, B. Fornberg,\, {Stable computation of differentiation matrices and scattered node stencils based on Gaussian radial basis functions},\,
SIAM J. Sci. Comput. 35 (4)(2013) 2096--2119.


\bibitem[Lee \emph{et al.}(1992)]{lee1993direct}
S. Lee, S.K. Lele, P. Moin,\, {Direct numerical simulation of isotropic turbulence interacting with a weak shock wave},\,
J. Fluid Mech. 251 (1993) 533--562.


\bibitem[Lele(1992)]{lele1992compact}
S.K. Lele,\, {Compact finite difference schemes with spectral-like resolution},\,
J. Comput. Phys. 103 (1)(1992) 16--42.


\bibitem[Liu \emph{et al.}(1998)]{liu1998preconditioned}
C. Liu, X.Zheng,  C. H. Sung,\, {Preconditioned multigrid methods for unsteady incompressible flows},\,
J. Comput. Phys. 139 (1)(1998) 35--57.



\bibitem[Moin(2010)]{moin2010fundamentals}
P. Moin,\,\textit{Fundamentals of engineering numerical analysis} (Cambridge University Press, 2010).


\bibitem[Nishikawa(2021)]{nishikawa2021flexible}
H. Nishikawa,\,{A flexible gradient method for unstructured-grid solvers},\, Internat. J. Numer. Methods Fluids, 93 (6)(2021) 2015--2021.






\bibitem[Patankar(2018)]{patankar2018numerical}
S. V. Patankar,\, {Numerical heat transfer and fluid flow},\,
CRC press. (2018).

\bibitem[Park and Mahesh \emph{et al.}(2007)]{park2007numerical}
N. Park, K. Mahesh,\, {Numerical and modeling issues in LES of compressible turbulence on unstructured grids},\,
Presented at AIAA Aerosp. Sci. Meet. Exhib., 45th, Reno, NV, AIAA Pap. (2007) 722.



\bibitem[Peng \emph{et al.}(2003)]{peng2003transition}
Y.-F. Peng, Y.-H. Shiau, R.R. Hwang,\, {Transition in a 2-D lid-driven cavity flow},\,
Comput. Fluids. 32 (3)(2003) 337--352.




\bibitem[Persson(2005)]{persson2005mesh}
P.O. Persson,\, Ph.D. thesis, Massachusetts Institute of Technology, 2005.


\bibitem[Powell(1992)]{powell1992theory}
M.J. Powell,\, {The theory of radial basis function approximation in 1990},\,
Adv. Numer. Anal. (1992) 105--210.



\bibitem[Rippa(1999)]{rippa1999algorithm}
S. Rippa,\, {An algorithm for selecting a good value for the parameter c in radial basis function interpolation},\,
Adv. Comput. Math. 11 (2)(1999) 193--210.




\bibitem[Russell and Wang(2003)]{russell2003cartesian}
 D. Russell, Z. J. Wang,\, {A Cartesian grid method for modeling multiple moving objects in 2D incompressible viscous flow},\,
J. Comput. Phys. 191 (1)(2003) 177--205.


\bibitem[Samtaney \emph{et al.}(2001)]{samtaney2001direct}
R. Samtaney, D.I. Pullin, B. Kosovi{\'c},\, {Direct numerical simulation of decaying compressible turbulence and shocklet statistics},\,
Phys. Fluids 13 (5)(2001) 1415--1430.



\bibitem[Santos \emph{et al.}(2018)]{santos2018comparing}
L.G.C. Santos, N. Manzanares-Filho, G.J. Menon, E. Abreu,\, {Comparing RBF-FD approximations based on stabilized Gaussians and on polyharmonic splines with polynomials},\,
Internat. J. Numer. Methods Engrg. 115 (4)(2018) 462--500.



\bibitem[Sanyasiraju and Chandhini(2008)]{sanyasiraju2008local}
 Y.V.S.S. Sanyasiraju, G. Chandhini,\, {Local radial basis function based gridfree scheme for unsteady incompressible viscous flows},\,
J. Comput. Phys. 227 (20)(2008) 8922--8948.


\bibitem[Sengupta \emph{et al.}(2011)]{sengupta2011analysis}
T.K. Sengupta, M.K. Rajpoot, S. Saurabh, V.V.S.N. Vijay,\, {Analysis of anisotropy of numerical wave solutions by high accuracy finite difference methods},\,
J. Comput. Phys. 230 (1)(2011) 27--60.

\bibitem[Shahane \emph{et al.}(2021)]{shahane2021high}
S. Shahane, A. Radhakrishnan, S.P. Vanka,\, {A high-order accurate meshless method for solution of incompressible fluid flow problems},\,
J. Comput. Phys. 445 (2021) 110623.


\bibitem[Shahane and Vanka(2021)]{shahane2021semi}
S. Shahane, S.P. Vanka,\, {A semi-implicit meshless method for incompressible flows in complex geometries},\,
arXiv preprint arXiv:2106.07616.


\bibitem[Shahane and Vanka(2022)]{shahane2022consistency}
S. Shahane, S.P. Vanka,\, {Consistency and Convergence of a High Order Accurate Meshless Method for Solution of Incompressible Fluid Flows},\,
arXiv preprint arXiv:2202.02828.


\bibitem[Shankar (2017)]{shankar2017overlapped}
V. Shankar,\, {The overlapped radial basis function-finite difference (RBF-FD) method: A generalization of RBF-FD},\,
J. Comput. Phys. 342 (2017) 211--228.

\bibitem[Shankar and Fogelson \emph{et al.}(2018)]{shankar2018hyperviscosity}
V. Shankar, A. L. Fogelson,\, {Hyperviscosity-based stabilization for radial basis function-finite difference (RBF-FD) discretizations of advection--diffusion equations},\,
J. Comput. Phys. 372 (2018) 616--639.



\bibitem[Shankar \emph{et al.}(2018)]{shankar2018rbf}
V. Shankar, A. Narayan, R. M. Kirby\, {RBF-LOI: Augmenting radial basis functions (RBFs) with least orthogonal interpolation (LOI) for solving PDEs on surfaces},\,
J. Comput. Phys. 373 (2018) 722--735.



\bibitem[Shu \emph{et al.}(2003)]{shu2003local}
C. Shu, H. Ding, K.S. Yeo,\, {Local radial basis function-based differential quadrature method and its application to solve two-dimensional incompressible Navier--Stokes equations},\,
Comput. Methods Appl. Mech. Engrg. 192 (7--8)(2003) 941--954.


\bibitem[Shu \emph{et al.}(2005)]{shu2005computation}
C. Shu, H. Ding, K.S. Yeo,\, {Computation of incompressible Navier-Stokes equations by local RBF-based differential quadrature method},\,
CMES Comput. Model. Eng. Sci. 7 (2)(2005) 195--206.



\bibitem[Su (2019)]{su2019radial}
L. Su,\, {A radial basis function (RBF)-finite difference (FD) method for the backward heat conduction problem},\,
Appl. Math. Comput. 354 (2019) 232--247.

\bibitem[Suzuki and Lele(2003)]{suzuki2003shock}
T. Suzuki, S.K. Lele,\, {Shock leakage through an unsteady vortex-laden mixing layer: application to jet screech},\,
J. Fluid Mech. 490 (2003) 139--167.


\bibitem[Tan \emph{et al.}(2021)]{tan2021two}
R. Tan, A. Ooi, R.D. Sandberg,\, {Two Dimensional Analysis of Hybrid Spectral/Finite Difference Schemes for Linearized Compressible Navier--Stokes Equations},\,
J. Sci. Comput. 87 (2)(2021) 1--41.



\bibitem[Tiesinga \emph{et al.}(2002)]{tiesinga2002bifurcation}
G. Tiesinga, F.W. Wubs, A.E.P. Veldman,\, {Bifurcation analysis of incompressible flow in a driven cavity by the Newton--Picard method},\,
J. Comput. Appl. Math. 140 (1-2)(2002) 751--772.






\bibitem[Tolstykh(2000)]{tolstykh2000using}
A.I. Tolstykh,\, {On using RBF-based differencing formulas for unstructured and mixed structured-unstructured grid calculations},\,
Proc. 16th IMACS World Congress. 228 (2000) 4606--4624.


\bibitem[Unnikrishnan \emph{et al.}(2022)]{unnikrishnan2021shear}
A. Unnikrishnan, S. Shahane, V. Narayan, S. P. Vanka,\, {Shear-driven flow in an elliptical enclosure generated by an inner rotating circular cylinder},\,
Phys. Fluids. 34 (1)(2022) 013607.




\bibitem[Wang and Wang(2002)]{wang2002point}
J.G. Wang, G. R. Wang,\, {A point interpolation meshless method based on radial basis functions},\,
Internat. J. Numer. Methods Engrg. 54 (11)(2002) 1623--1648.



\bibitem[Wang(2017)]{wang2017radial}
L. Wang,\, {Radial basis functions methods for boundary value problems: Performance comparison},\,
Eng. Anal. Bound. Elem. 84 (2017) 191--205.


\bibitem[Wang \emph{et al.}(2020)]{wang2020weighted}
L. Wang, Z. Qian, Y. Zhou, Y. Peng,\, {A weighted meshfree collocation method for incompressible flows using radial basis functions},\,
J. Comput. Phys. 401 (2020) 108964.


\bibitem[Wang \emph{et al.}(2021)]{wang2021static}
L. Wang, Y. Liu, Y. Zhou, Y. Fan,\, {Static and dynamic analysis of thin functionally graded shell with in-plane material inhomogeneity},\,
Int. J. Mech. Sci. 193 (2021) 106165.

\bibitem[Waters and Pepper(2015)]{waters2015global}
J. Waters, D. W. Pepper,\, {Global versus localized RBF meshless methods for solving incompressible fluid flow with heat transfer},\,
Numer. Heat Transf. B: Fundam. 68 (3)(2015) 185--203.



\bibitem[Wright (2003)]{wright2003radial}
G.B. Wright,\, {Radial basis function interpolation: numerical and analytical developments},\,
PhD thesis, University of Colorado.


\bibitem[Wright \emph{et al.}(2010)]{wright2010hybrid}
G.B. Wright, N. Flyer, D. A. Yuen,\, {A hybrid radial basis function--pseudospectral method for thermal convection in a 3-D spherical shell},\,
Geophys. Geochem. Geosyst. 11 (7)(2010).

\bibitem[Wright and Fornberg(2006)]{wright2006scattered}
G.B. Wright, B. Fornberg,\, {Scattered node compact finite difference-type formulas generated from radial basis functions},\,
J. Comput. Phys. 212 (1)(2006) 99--123.





\bibitem[Xie \emph{et al.}(2021)]{xie2021improved}
Y. Xie, X. Zhao, M. Rubinato, Y. Yu,\, {An improved meshfree scheme based on radial basis functions for solving incompressible Navier-Stokes equations},\,
Int. J. Numer. Meth. Fluids. 93 (2021) 2842--2862.


\bibitem[Zamolo and Nobile(2019)]{zamolo2019solution}
R. Zamolo, E. Nobile,\, {Solution of incompressible fluid flow problems with heat transfer by means of an efficient RBF-FD meshless approach},\,
Numer. Heat Transf. B: Fundam. 75 (1)(2019) 19--42.





\end{thebibliography}

\end{document}